\documentclass[british]{iopart}

\usepackage{iopams}
\usepackage{babel,setstack}
\usepackage{graphicx}
\usepackage{amsthm}
\usepackage{color}
\usepackage[sans]{dsfont}
\usepackage[mathscr]{euscript}

\usepackage{cite}
\usepackage{times}

\newtheorem{assumption}{Assumption}

\newcommand{\envS}{\sigma_{\text{env}}}
\newcommand{\ranenv}{\rangle_{\text{env}}}
\newcommand{\rmm}{\mathrm{m}}
\newcommand{\openone}{\mathds{1}}

\newcommand{\RE}{\mathrm{\,Re\,}}
\newcommand{\IM}{\mathrm{\,Im\,}}
\newcommand{\Ebb}{\mathbb{E}}
\newcommand{\Rbb}{\mathbb{R}}
\newcommand{\Cbb}{\mathbb{C}}

\newcommand{\om}{{\omega_{\rm m}}}
\newcommand{\oc}{{\omega_{\rm c}}}
\newcommand{\omq}{{\omega_{\rm m}^{\,2}}}
\newcommand{\Om}{{\Omega_{\rm m}}}
\newcommand{\Omq}{{\Omega_{\rm m}^{\,2}}}
\newcommand{\cgamq}{{\gamma_{\rm c}^{\,2}}}
\newcommand{\cgam}{{\gamma_{\rm c}}}
\newcommand{\mgamq}{{\gamma_{\rm m}^{\,2}}}
\newcommand{\mgam}{{\gamma_{\rm m}}}

\newcommand{\norm}[1]{\left\Vert#1\right\Vert}
\newcommand{\abs}[1]{\left\vert#1\right\vert}

\newcommand{\Ccal}{\mathcal{C}}
\newcommand{\Lcal}{\mathcal{L}}

\newcommand{\Wcal}{\mathcal{W}}

\begin{document}

\title{Quantum Langevin equations for optomechanical systems}

\author{Alberto Barchielli$^{1,3,4}$ and Bassano Vacchini$^{2,3}$}
\address{$^1$ Politecnico di Milano, Dipartimento di Matematica,
piazza Leonardo da Vinci 32, 20133 Milano, Italy}
\address{$^2$ Universit{\`a} degli Studi di Milano, Dipartimento di Fisica,
Via Celoria 16, 20133 Milano, Italy}
\address{$^3$ Istituto Nazionale di Fisica Nucleare (INFN), Sezione di Milano,}
\address{$^4$ Istituto Nazionale di Alta Matematica (INDAM-GNAMPA)}

\eads{\mailto{alberto.barchielli@polimi.it}, \
\mailto{bassano.vacchini@mi.infn.it}}
\begin{abstract}
We provide a fully quantum description of a mechanical oscillator in the
presence of thermal environmental noise by means of a quantum Langevin
formulation based on quantum stochastic calculus. The system dynamics is
determined by symmetry requirements and equipartition at equilibrium, while the
environment is described by quantum Bose fields in a suitable non-Fock
representation which allows for the introduction of temperature. A generic
spectral density of the environment can be described by introducing its state
through a suitable $P$-representation. Including interaction of the mechanical
oscillator with a cavity mode via radiation pressure we obtain a description of
a simple optomechanical system in which, besides the Langevin equations for the
system, one has the exact input-output relations for the quantum noises. The
whole theory is valid at arbitrarily low temperature. This allows the exact
calculation of the stationary value of the mean energy of the mechanical
oscillator, as well as both homodyne and heterodyne spectra. The present
analysis allows in particular to study possible cooling scenarios and to obtain
the exact connection between observed spectra and fluctuation spectra of the
position of the mechanical oscillator.
\end{abstract}


\submitto{\NJP}

\tableofcontents

\section{Introduction}
Optomechanical systems in the quantum regime are very important for quantum
information processing and for testing fundamental issues of quantum mechanics
\cite{Bahrami2014a,GMVT09,SDHB12,LGP13,PGKBBA06,SN-P13,Chen13,Pontin14,asp14,JJ14}.
Their theoretical analysis therefore calls for a first principle description.
In particular since the focus is on quantum effects, the theoretical models
must be fully consistent with quantum mechanics. Actually the correct quantum
description of a mesoscopic mechanical oscillator and of the thermal noise
affecting it is not a trivial task, and there is not a unique accepted model
for them \cite{Lin76b,Dek81,Caldeira81,Caldeira1983b,Dio93a,Dio93b,
FLC1,FLC2,SanS87,Va00,Va02a,XZWW10,ZLXTN12,YCXZS13}.

The first aim of this paper is therefore to obtain an accurate quantum
mechanical description of a mechanical oscillator taken to be part of an
opto-mechanical device. The oscillator cannot be considered as a Brownian
particle, but rather as a mesoscopic mechanical system, say a movable mirror
mounted on a vibrating structure. Dissipative effects are essentially due to
the interaction with phonons. Our strategy will be to introduce reasonable
physical requirements leading to a master equation in Lindblad form, valid for
any temperature of the thermal bath. We then translate these results into
quantum Langevin equations and we show how to obtain a suitable non-Markovian
generalization at this level of description. Relying on these results we can
consider the description of the simplest optomechanical system, that is a
moving mirror interacting with an electromagnetic mode in a cavity via
radiation pressure \cite{PGKBBA06,Genes07,SN-P13,Chen13,Pontin14}. Again a
suitable analysis of the composite system and of the monitoring of the emitted
light calls for a consistent quantum description. We shall obtain this result
by the use of quantum Langevin equations, directly deducing them from a unitary
dynamics, and exploiting the theory of measurements in continuous time.

The paper is organized as follows. In \Sref{sec:MELin} we determine the reduced
dynamics of the mechanical oscillator. Here, the basic assumption is the use of
a Markovian master equation with a quadratic generator and having a unique
equilibrium state. Its structure is further determined by suitable symmetry
requirements and by physical constraints on the behaviour of the mean values of
position and momentum. In \Sref{Lang} we introduce the quantum Langevin
equations for the mechanical oscillator alone; the whole presentation is based
on the notions of quantum noise \cite{GarZ00,Car08} and of input-output fields
\cite{GarC85,Bar86,Bar06}, as well as on the use of quantum stochastic calculus
\cite{HudP84,Parthas92}. The Bose fields entering in the unitary dynamics play
the role of phonon fields. By modifying their state without changing the time
evolution operator it is possible to introduce non Markovian effects, namely a
non-flat noise spectrum. The differences with respect to usual approaches are
of relevance especially at low temperatures, where the zero-point fluctuations
play an essential role.

A quantum optomechanical system is studied in \Sref{sec:model1} by using the
quantum Langevin approach, within a fully consistent formulation valid at any
temperature. Firstly, the typical effect of laser cooling is discussed. Then,
the continuous monitoring of the emitted light is introduced in Sections
\ref{sec:homo} (homodyne detection) and \ref{sec:hetero} (heterodyne
detection). The treatment is well based in the theory of measurements in
continuous time. Detection of the emitted light is usually assumed to give a
direct measurement of the fluctuations of the position of the mechanical
component. We show that this is true, but only for not too low temperatures; at
very small temperatures, interference terms are important and the direct
connection with such fluctuations is lost. This leads to new predictions on the
optical spectra at very low temperature. We finally summarize and discuss our
results in \Sref{sec:concl}.

\section{Damped mechanical oscillator: the master equation approach}\label{sec:MELin}
As a first step towards the construction of models of optomechanical systems
valid in the quantum regime at low temperatures, we consider the reduced
dynamics of an open mechanical oscillator. A fully consistent quantum
description of a massive nanomechanical component, kept at  the simplest
possible level, will be our basic building block in order to consider more
complex dynamics. We therefore formulate in the first instance a Markovian
description for the mechanical oscillator, which we build up relying on general
physical constraints and symmetry requirements.

A standard approach often considered in the literature is to derive the master
equation for the harmonic oscillator from effective environmental models of
bosonic oscillators
\cite{Caldeira81,Caldeira1983b,FLC1,FLC2,XZWW10,ZLXTN12,YCXZS13}. However,
previous work \cite{Dio93a,Dio93b} has shown that, while using careful
approximations a positive Markovian dynamics can be obtained in this framework,
the final results are valid only from medium to high temperatures of the
thermal bath. To ask for a Markovian dynamics based on symmetry arguments
allows to get simpler and more universal models, but again serious problems
appear. The requirement that the equilibrium state should be the canonical
thermal state determined by the standard Hamiltonian of a harmonic oscillator
is known to be incompatible with positivity and translational invariance
\cite{Lin76b,Tannor97,PeixN98}. This incompatibility induced some authors to
renounce to translational invariance \cite{SanS87,PeixN98}, or to accept
non-positive dynamical equations and to give more relevance to obtaining time
evolutions very close to the classical ones
\cite{Caldeira81,Caldeira1983b,DioGTY99}. A non positive dynamics can be
satisfactory when the system is near the classical regime, but this approach
becomes questionable when quantum effects are searched for
\cite{GioV01,JacTWS99}.  We shall show that it is possible to maintain
positivity and translational invariance by weakening the requests on the
equilibrium state. The key point will be a weak formulation of energy
equipartition at equilibrium. Our result is therefore to single out from the
many proposals appearing in the literature a unique consistent dynamics in the
Markov approximation.

\subsection{Physical constraints and symmetry requirements}
We formulate now our assumptions, starting from the existence of a well defined
positive Markovian dynamics, describing damping and translationally invariant
apart from the harmonic potential. A weak equipartition condition and the
existence of a unique stationary state in Gibbs form, as we shall see, will
essentially fix the structure of the reduced dynamics.

\begin{assumption}[Positive Markovian dynamics with quadratic
  generator] The evolution of the statistical operator
  of the oscillator is governed by a Markovian master equation
  preserving the positivity of the states. The generator of the
  dynamics is at most quadratic in the position and momentum operators
  of the mechanical oscillator.\end{assumption}

\noindent The first assumption is to consider a time-homogeneous and linear
time evolution. Such a dynamics can be expressed in the form
\begin{equation}
\frac{\rmd \ }{\rmd t}\,\rho(t)=\Lcal[\rho(t)],
\end{equation}
with $\Lcal$ a suitable generator or Liouville operator, at most quadratic in
the position and momentum operators of the mechanical oscillator $q$ and $p$,
so as to have at most a quadratic potential term and a friction effect
proportional to the momentum of the mechanical oscillator. In the case of
linear systems it is known that positivity and complete positivity of the
dynamics are actually equivalent \cite{Lin76b}, therefore according to
\cite{GorKS76,Lin76} the generator $\Lcal$ must have the standard Lindblad
structure. The most general quadratic Liouville operator is obtained in terms
of  two Lindblad operators \cite{Lin76b}
\begin{equation}\label{R_j}
R_j=\frac 1{\sqrt \hbar}\left(u_j q +v_j p\right),\qquad u, v\in \Cbb^2,
\end{equation}
and a generic selfadjoint quadratic Hamiltonian for the mechanical system
\begin{equation*}
H_\rmm=\frac {h_q} 2\,q^2 + \frac {\kappa_0}  4\left\{q,p\right\}+ \frac {h_p} 2\, p^2+f_qq+f_pp,
\end{equation*}
where all the constants are taken to be real, so that $\Lcal$ takes the form
\begin{equation}
\Lcal[\rho]=-\frac\rmi\hbar \left[H_\rmm,\rho\right]+\sum_{j=1}^2\left(R_j\rho
R_j^{\,\dagger}-\frac 1 2 \left\{R_j^{\,\dagger}R_j,\rho\right\}\right).
\end{equation}

We ask now to have a damped oscillator, but not an \emph{overdamped} one.
\begin{assumption}[Damping] The ``kinetic energy''
term is non negative and the mean values of position and momentum decay to zero
with an oscillating behaviour.\end{assumption}
 \noindent Apart from the trivial requirement of a positive
\emph{kinetic energy} term, we further look for a dynamics describing the
oscillating decay of the mean values of $q$ and $p$ to zero. This condition
complies with the Markovian and quadratic approximations, which are expected to
be good only for small damping. Denoting by $\langle X\rangle_t$ the mean value
of a quantum operator with the state $\rho(t)$ solution of the master equation
we have for position and momentum
\begin{eqnarray*}
\frac{\rmd \langle q \rangle_t}{\rmd t}&=h_p \langle p \rangle_t+
\left(\frac {\kappa_0} 2-\IM\langle u|v\rangle\right)\langle q \rangle_t+f_p,
\\
\frac{\rmd \langle  p\rangle_t }{\rmd t}&=-h_q\langle q\rangle_t-
\left(\frac {\kappa_0} 2+\IM\langle u|v\rangle\right)\langle p\rangle_t -f_q,
\end{eqnarray*}
with $\langle u|v\rangle$ the inner product in $\Cbb^2$. The eigenvalues of the
associated dynamical matrix are $-\IM\langle u|v\rangle \pm \sqrt{{\kappa_0
^2}/4-h_ph_q}$, so that in order to have an underdamped dynamics we need
$\IM\langle u|v\rangle>0$ and ${\kappa_0 ^2}/4<h_ph_q$. In particular $h_p$ and
$h_q$ have the same sign and are non-vanishing. Positivity of the kinetic
energy leads to $h_p>0$ and therefore $h_q>0$. Then, we can write $h_p=1/m$ and
$h_q=m\Omq$; the above inequality on $\kappa_0$ becomes $\kappa_0^{\,2}<4\Omq$.
Finally, the vanishing of the equilibrium means imply $f_q=0$, $f_p=0$.
Introducing the positive coefficients
\begin{equation*}
\fl
\mgam =2\IM\langle u|v\rangle, \qquad D_{qp}=\RE\langle u|v\rangle,
\qquad D_{qq}=\norm v^2, \qquad D_{pp}=\norm u^2,
\end{equation*}
the generator can be written in the form
\begin{eqnarray}\label{nuova}\nonumber
\fl \hbar\Lcal[\rho]=-\frac{\rmi }{2m}\left[p,\left\{p,\rho\right\}\right]
-\frac{\rmi m\Omq }2\left[q,\left\{q,\rho\right\}\right]-\frac {D_{pp}}2\,[q,[q,\rho]]
-\frac{D_{qq}}2\,[p,[p,\rho]]\\ {} -D_{qp}\, [p,[q,\rho]]
- \frac {\rmi\left(\kappa_0 +\mgam \right)} 4\left[q,\left\{p,\rho\right\}\right]
-\frac {\rmi\left(\kappa_0 -\mgam  \right)}4\left[p,\left\{q,\rho\right\}\right], \label{Liouv1}
\end{eqnarray}
where in particular the constraints
\begin{equation}\label{ineq3}
D_{qq}\geq 0, \qquad D_{pp}\geq 0, \qquad D_{qq}D_{pp}-D_{qp}^{\;2}
-\left(\frac {\mgam  } 2\right)^2\geq 0
\end{equation}
hold, which provide the necessary and sufficient conditions for the dynamics
described by \eref{nuova} to be in Lindblad form and therefore completely
positive \cite{Lin76,Lin76b}. An alternative way to get the same positivity
condition is to ask the \emph{generalized} Heisenberg uncertainty relation
$\langle q^2\rangle_t\langle p^2\rangle_t-({\langle\{p,q\}\rangle_t}/2)^2\geq
{\hbar^2}/4$ to hold for any time and any initial state \cite{DekV84}.

The first two assumptions are implicitly or explicitly taken in all the
Markovian approaches. A further natural requirement is that the interaction
with the environment does not depend on the position of the oscillator.

\begin{assumption}[Translational invariance] The reduced dynamics is
invariant under translations apart from the harmonic potential term. This
requirement is equivalent to the validity of the classical equations of motion
for the mean values of position and momentum.
\end{assumption}
\noindent By applying the generic translation $q \mapsto q+x$, $p\mapsto p$ to
the generator \eref{nuova} we see that all the terms are invariant with the
exclusion of the term containing the harmonic potential $-\frac{\rmi m\Omq
}2\left[q,\left\{q,\rho\right\}\right]$ and the last term $- \frac
{\rmi\left(\kappa_0 -\mgam \right)}4\left[p,\left\{q,\rho\right\}\right]$.
Therefore, the above assumption is satisfied if and only if $\kappa_0 = \mgam
$. The same conclusion is reached by considering the equations of motion for
position and momentum and asking them to be equivalent to the classical
equations in which the momentum is proportional to the derivative of the
position.

The result of the first three assumptions is therefore that the Liouville
operator has the structure \eref{nuova} with $\kappa_0 = \mgam
>0$ and $\Omq >
{\mgamq}/4$; moreover, the constraints \eref{ineq3} hold. In particular, the
Hamiltonian part of $\Lcal$ turns out to be
\begin{equation}\label{H0H00}
H_\rmm=H_0 + \frac {\mgam}  4\left\{q,p\right\},
\qquad H_0=\frac {p^2}{2m} + \frac 1 2\,m\Omq q^2,
\end{equation}
where, besides a contribution in the form of the free Hamiltonian of a harmonic
oscillator with a strictly positive frequency $\Om$, one has an additional term
in the form of an anticommutator proportional to the damping constant.

The evolution equations for the mean values and second moments of position and
momentum then read:
\begin{equation}\label{eq_for_q,p}
\frac{\rmd \langle q\rangle_t}{\rmd t}=\frac {\langle p\rangle_t} m ,
\qquad
\frac{\rmd \langle p\rangle_t}{\rmd t}=- m\Omq \langle q\rangle_t-\mgam \langle p\rangle_t,
\end{equation}
\begin{equation}\label{eq:q2.}\eqalign{
\frac{\rmd \langle q^2\rangle_t}{\rmd t}=\frac{\langle\left\{p,q\right\}\rangle_t}m+\hbar D_{qq},
\cr
\frac{\rmd \langle p^2\rangle_t}{\rmd t}=-m\Omq\langle\left\{p,q\right\}\rangle_t
- 2\mgam \langle p^2\rangle_t+\hbar D_{pp},
\cr
\frac{\rmd \langle \{q,p\}\rangle_t}{\rmd t}=\frac{2\langle p^2\rangle_t}m -
2m\Omq \langle q^2\rangle_t -\mgam \langle \{q,p\} \rangle_t -2\hbar D_{qp}.}
\end{equation}
The dynamical matrix giving the evolution of the mean values \eref{eq_for_q,p}
has eigenvalues $-{\mgam }/2$ and $-{\mgam }/2\pm \rmi \sqrt{\Omq-{\mgamq}/4}$,
which naturally leads to introduce the \emph{damped} frequency $\om$ of the
mechanical oscillator in terms of its \emph{bare} frequency $\Om$:
\begin{equation}
\om=\sqrt{\Omq-\frac{\mgamq }4}.
\end{equation}
Let us note that according to $\Omq>\mgamq /4>0$ we have ruled out the case
$\Om=0$, which corresponds to a quantum Brownian particle, that is a massive
particle not bounded by a potential in a translation invariant environment
\cite{Diosi1995a,Va00,Va01a,VaP05,Hornberger2006b,Hornberger2008a} (see
\cite{VaH09} for a recent review).

At this stage we further have to determine the diffusion coefficients $D_{qp}$,
$D_{qq}$ and $D_{pp}$ appearing in \eref{Liouv1}. We will rely on the study of
features of the equilibrium state, but to avoid the known incompatibilities
with translation invariance \cite{Lin76b} we formulate the equipartition
condition in a weaker form, not asking the existence of an equilibrium Gibbs
state generated by $H_{0}$.
\begin{assumption}[Equipartition] At equilibrium the mean kinetic
  energy and the mean potential energy  have to be equal.
\end{assumption}
\noindent Since the eigenvalues of the dynamical matrix associated to
\eref{eq:q2.} have a positive real part, \emph{existence of a unique attractive
equilibrium state} is granted, and thanks to the linearity of the equations the
equilibrium state is actually \emph{Gaussian} and determined by the mean values
at equilibrium. Then, our \emph{equipartition} condition is
\begin{equation}\label{equip}
\frac{\langle p^2 \rangle_{\mathrm{eq}}}{2m}=\frac{1}2 \,m \Omq\langle q^2 \rangle_{\mathrm{eq}},
\end{equation}
which gives equal weight to the mean kinetic and potential energy at
equilibrium. By setting in \eref{eq:q2.} the time derivatives equal to zero we
come to
\begin{equation}\label{eq:equi}
D_{qp}=\frac {m\mgam } {2}\,D_{qq}.
\end{equation}
Moreover, the second moments at equilibrium turn out to be
\begin{equation}\label{equi3}
\eqalign{
\langle q^2 \rangle_{\mathrm{eq}}=\frac {\hbar }{2\mgam }
\left(D_{qq}+\frac{D_{pp}}{m^2\Omq} \right), \qquad \langle \{q,p\}
\rangle_{\mathrm{eq}}=-\hbar m D_{qq},
\cr
\langle p^2 \rangle_{\mathrm{eq}}=\frac {\hbar }{2\mgam }
\left(D_{pp}+m^2\Omq D_{qq}\right).}
\end{equation}

We exploit finally the residual freedom we have in the choice of the diffusion
coefficients to get a Gibbs state as equilibrium state. However, as we already
noticed, it cannot be state generated by $H_{0}$, and we replace it by a
generic effective Hamiltonian.
\begin{assumption}[Gibbs state and temperature dependence]\label{as:5}
The equilibrium state has a Gibbs form which is determined by an effective
Hamiltonian independent from the temperature.
\end{assumption}
\noindent As we shall prove below, this assumption implies that the diffusion
coefficients have the expressions
\begin{equation}\label{finalD}
\fl
D_{qq}=\frac{\mgam \left(2N+1\right)}{2m\om}, \qquad D_{pp}=\frac{\mgam  m\Omq}{2\om}\left(2N+1\right),
\qquad D_{qp}=\frac{\mgamq }{4\om}\left(2N+1\right)
\end{equation}
and that the equilibrium state has the form
\begin{equation}\label{eqstate}
\rho_{\mathrm{eq}}= \frac{\rme^{- \beta H_\rmm}}{\Tr\left\{\rme^{-\beta H_\rmm}\right\}}\,.
\end{equation}
In these expressions, $H_\rmm$ is the mechanical Hamiltonian \eref{H0H00} and
$N\geq0$ represents the \emph{mean number of excitations} in the equilibrium
state, namely
\begin{equation}\label{def:N} N=\frac 1
{\rme^{\beta\hbar\om}-1}.
\end{equation}

Let us prove the above statements. Thanks to the Gaussianity of the equilibrium
state,  Assumption \ref{as:5} means $\rho_{\rm eq} \propto \exp\left\{- c
\tilde H\right\}$ for a suitable quadratic Hamiltonian (without the liner
terms, because the means of position and momentum have to be zero), say $\tilde
H=\frac {p^2}{2\tilde m} + \frac 1 2\,\tilde m\tilde\Omega^2+ \frac {\tilde
\gamma^2} 4\left\{q,p\right\}$, with $\tilde \omega^2:=
\tilde\Omega^2-\frac{\tilde\gamma^2}4>0$ in order $\rho_{\rm eq}$ to be a
trace-class operator. Then, the eigenvalues of $\tilde H$ have the form $\hbar
\tilde \omega\left(n+1/2\right)$ and, without changing $\rho_{\rm eq}$, we can
redefine $c$ and $\tilde H$ in such a way that $\tilde \omega=\om$ and
$c=\beta$, a positive constant which can be interpreted as the inverse
temperature of the equilibrium state of the mechanical oscillator. Then, the
mean number of excitations has the expression \eref{def:N}. By equating the
mean values determined by $\rho_{\rm eq}$ with the expressions \eref{equi3},
after some algebraic manipulations we get
\[
 \tilde m^2\tilde \Omega^2=m^2\Omq, \qquad \frac{m\mgam}{\tilde m \tilde \gamma}
 =\frac{mD_{qq}+\frac{D_{pp}}{m\Omq}}{2mD_{qq}},
\]
\[
\left(2N+1\right)^2=\frac\Omq\mgamq\left(\frac{D_{pp}}{m\Omq}-mD_{qq}\right)^2
+\frac 4 \mgamq \left(D_{qq}D_{pp}-\frac {m^2\mgamq}4\, D_{qq}^{\;2}\right).
\]
The right hand side of the last equation is greater or equal to 1 by
\eref{ineq3} and \eref{eq:equi}. To have $\tilde H$ independent from the
temperature implies that the coefficients $D_{qq}$ and $D_{pp}$ are both
proportional to $2N+1$ with temperature independent coefficients. The equations
above, together with $\tilde \omega=\om$, give the expressions \eref{finalD}
and $\tilde H=H_\rmm$.

\subsection{Master equation for the mechanical oscillator}\label{finalME}
From the previous results we see that a central role is played by the
mechanical Hamiltonian $H_\rmm$, which appears in the commutator part of the
Liouville operator and determines the equilibrium state \eref{eqstate}. It will
be very useful to diagonalize explicitly $H_\rmm$ by introducing a suitable
mode operator. By defining
\begin{equation}\label{atau}
a_\rmm   = \frac{1}{\sqrt{2 m\hbar\om}} \left( m\Om \,q +  {\rmi \tau}\, p\right),
\qquad \tau=\frac{\om}{\Om}-\frac{\rmi}{2} \frac{\mgam }{\Om},
\end{equation}
we get that the mechanical Hamiltonian \eref{H0H00} can be written as
\begin{equation}\label{Hm}
H_\rmm
=\hbar \om \left(a_\rmm ^\dagger a_\rmm  +\frac 1 2 \right).
\end{equation}
The dimensionless quantity $\tau$ has modulus equal to one and
$[a_\rmm,a_\rmm^\dagger]=1$ is satisfied. The inverse formulae giving $q,\,p$
in terms of $a_\rmm,\,a_\rmm^\dagger$ are
\begin{equation}\label{qp-a}
q=\sqrt{\frac\hbar{2m\om}}\left(\overline \tau \, a_\rmm  +  \tau a_\rmm ^\dagger\right),
\qquad p=\rmi\sqrt{\frac{m\hbar \Omq}{2\om}}\left(a_\rmm ^\dagger - a_\rmm \right).
\end{equation}

Let us note that the form \eref{eqstate} of the equilibrium state does not come
from a direct requirement, but rather it follows from all the considered
assumptions. In particular we stress the fact that the operator $H_\rmm$ is not
the Hamiltonian of the isolated oscillator, but includes a term containing
$\mgam $ which comes from the interaction with the bath. Combining \eref{equi3}
and \eref{finalD} we have in particular
\begin{equation}\label{eqvalues}
\fl
\frac{\langle p^2 \rangle_{\mathrm{eq}}}{2m}=\frac 12\,m\Omq \,\langle q^2
\rangle_{\mathrm{eq}}=\frac {\hbar \Omq }{4\om}\left(2N+1\right),  \qquad
\frac \mgam  4\langle \{q,p\} \rangle_{\mathrm{eq}}=- \frac {\hbar\mgamq } {8\om} \left(2N+1\right),
\end{equation}
indeed satisfying \eref{ineq3}. We see that the term related to damping gives a
negative contribution to  the equilibrium mean value $\langle H_\rmm
\rangle_{\mathrm{eq}}$ arising from energy exchange with the bath. The Lindblad
operators $R_j$ appearing in \eref{R_j} now read $R_1=\sqrt{\mgam
\left(N+1\right)} \, a_\rmm$, $R_2=\sqrt{\mgam  N} \, a_\rmm ^\dagger$ so that
the Liouville operator can be finally written as
\begin{eqnarray}\nonumber\fl
\Lcal[\rho]=-\frac{\rmi}{\hbar} \left[ H_\rmm ,\rho\right]+\mgam  (N+1) \left( a_\rmm \rho a_\rmm ^\dagger
- \frac 12 \left\{a_\rmm ^\dagger a_\rmm ,\rho\right\}\right)
\\ {}+\mgam  N\left( a_\rmm^\dagger \rho a_\rmm
- \frac 12 \left\{a_\rmm a_\rmm ^\dagger ,\rho\right\}\right).\label{Lcala}
\end{eqnarray}
Let us stress that, despite the fact that the expression \eref{Lcala} has the
form of the generator for an optical oscillator \cite{GarZ00}, the relations
\eref{atau}, \eref{qp-a} connecting $a_\rmm ,\,a_\rmm ^\dagger$ with $q,\,p$
account for the description of a mechanical oscillator. A master equation for a
mechanical oscillator with the Liouville operator \eref{Lcala} and the relation
\eref{atau} between mode and position/momentum operators was already proposed
in \cite[Sects.\ 6, 7]{Dek81}, inside a scheme of canonical quantization of
dissipative classical systems. The introduction of \eref{atau} in order to
obtain a quantum description of the mechanical oscillator complying with all
the natural physical requirements is a key result of this section, which we
will later exploit to consistently treat optomechanical systems.

\section{Langevin equations for the mechanical oscillator}\label{Lang}

So far we have obtained a quantum master equation in Lindblad form for the
mechanical oscillator, only relying on general physical constraints and
symmetry requirements. However, optomechanical systems are typically dealt with
making use of quantum Langevin equations, which provide a suitable and powerful
approach for linear systems \cite{GioV01}; in such a framework not only the
system of interest appears, but also some quantum noises representing the
environment. It is a general result that for any master equation in Lindblad
form it is possible to introduce in a rigourous way a unitary dynamics
involving the system of interest and suitable quantum Bose fields, which at the
level of the reduced dynamics of the system exactly reproduces the master
equation. That is, these quantum Bose fields effectively describe the thermal
environment affecting the mechanical oscillator and the system/field dynamics
is given by a unitary time evolution operator satisfying a quantum stochastic
differential equation of the type introduced by Hudson and Parthasarathy
\cite{HudP84}. Within this formalism the Heisenberg equations for the system
operators provide the quantum Langevin equations, while, as shown in
\cite{Bar86}, the Heisenberg equations for the Bose fields give the
input-output relation of Gardiner and Collet \cite{GarC85,GarZ00}.  We thus
obtain in a unified framework all relevant physical information \cite{Bar06}.
Finally, we shall show in \Sref{sec:n-M} that this approach allows to treat
also non Markovian effects and to introduce noises with non flat spectrum.

Let us start introducing the Hudson-Parthasarathy equation or quantum
stochastic Schr\"odinger equation \cite{HudP84}, which gives the evolution
equation for the unitary dynamics involving the system of interest and a
quantum Bose field. The proper mathematical formulation of this equation relies
on the formalism of quantum stochastic calculus \cite{Parthas92}. For the
Liouville operator \eref{Lcala} the associated Hudson-Parthasarathy equation
reads (see e.g.\ \cite{LinW86,Bar86,AttJ07} or \cite[Sections 11.2.2,
11.2.7]{GarZ00})
\begin{eqnarray}\fl\nonumber
\rmd U(t)=\biggl\{-\frac\rmi \hbar\,H_\rmm\rmd t + \sqrt \mgam  \left (a_\rmm\rmd
B_{\rm th}^\dagger(t) -a_\rmm^\dagger\rmd
B_{\rm th}(t)\right)
\\
{}-\frac \mgam {2 }\left(\left(2N+1\right) a_\rmm^\dagger
  a_\rmm+N\right)\rmd t \biggr\}U(t),
\label{eq:U}
\end{eqnarray}
with $U(0)=\openone$, $H_\rmm$ given by \eref{Hm}, and $B_{\rm th}(t)$ a Bose
thermal field satisfying the canonical commutation relations
\begin{equation}\label{eq:3}
[B_{\rm th}(t),B_{\rm th}^\dagger(s)]
= {\min\{t,s\}}$, \qquad $[B_{\rm th}(t),B_{\rm th}(s)]=0,
\end{equation}
and the \emph{quantum It\^o table}
\begin{equation}\label{Ito}\eqalign{
\rmd B_{\rm th}(t)\,\rmd B_{\rm th}^\dagger(t)= (N+1)\,\rmd t,\qquad
\rmd B_{\rm th}^\dagger(t)\,\rmd B_{\rm th}(t)= N\,\rmd t,
\cr \rmd B_{\rm th}(t)\,\rmd B_{\rm th}(t)= 0, \qquad \qquad
\rmd B_{\rm th}(t)\rmd t=\rmd B_{\rm th}^\dagger(t)\rmd t= 0,}
\end{equation}
with $N$ the positive quantity introduced in \eref{def:N}. The commutation
rules are better understood by introducing the formal field densities: $\rmd
B_{\rm th}(t) =b_{\rm th}(t)\, \rmd t$. Then, these densities satisfy the
standard canonical commutation relations
\begin{equation}\label{deltacomm}
[b_{\rm th}(t),b_{\rm th}^\dagger(s)]=\delta(t-s),\qquad [b_{\rm th}(t),b_{\rm th}(s)]=0.
\end{equation}

Equation \eref{eq:U} is a quantum stochastic differential equation in It\^o
sense and the second line of \eref{eq:U} corresponds to the \emph{It\^o
correction}. The solution $U(t)$ of \eref{eq:U} is a family of unitary
operators on the overall Hilbert space which represent the dynamics of the
closed system corresponding to ``mechanical oscillator plus field''. An
heuristic, but more familiar, picture can be obtained by using the field
densities introduced above. The formal expression of the unitary evolution is
indeed \cite{BarG13}
\begin{equation}\label{TU(t)}
U(t)= \overleftarrow{\mathrm {T}} \, \exp \Bigl\{  \int_0^t
\left[-\frac{\rmi H_\rmm} \hbar+ \sqrt \mgam \left( a_\rmm b_{\rm th}^\dagger(s)-a_\rmm^\dagger
b_{\rm th}(s)\right) \right] \rmd s \Bigr\},
\end{equation}
where $\overleftarrow{\mathrm {T}} $ denotes the time ordered product. From
this formal expression one sees that $U(t)$ is the time evolution operator for
system and field in the interaction picture with respect to the free field
dynamics. The thermal field $B_{\rm th}$ therefore represents the environment,
say the phonon field interacting with the mechanical oscillator.

It is possible to show that the physical thermal field $B_{\rm th}(t)$ does not
admit a Fock representation. However, it is useful for computations to have a
hand a mathematical representation of $B_{\rm th}(t)$ in terms of two commuting
Bose fields $A_1$ and $A_2$ in the Fock representation \cite{Bar86,AttJ07}.
This means that such fields satisfy the canonical commutations rules
$[A_i(t),A_j^\dagger(s)]=\delta_{ij} {\min\{t,s\}}$, $[A_i(t),A_j(s)]=0$  and
that there exists a common Fock vacuum, i.e.\ a normalized vector $e(0)$
annihilated by all these operators: $A_k(t)e(0)=0$ for $k=1,2$. The field
defined by
\begin{equation}\label{thfield}
B_{\rm th}(t)=\sqrt{N+1}\,   A_1(t) -\sqrt{N} \,   A^\dagger_2(t),
\end{equation}
satisfies the canonical commutation relations \eref{eq:3} and the It\^o table
\eref{Ito}. It is known in quantum field theory that there exist non unitarily
equivalent representations of the canonical commutation relations; indeed,
$B_{\rm th}(t)$ cannot be obtained by unitary transformations of Fock fields.

Let us now consider as state of the field the $A$-field vacuum $e(0)$. In such
a case taking the partial trace over the Fock space of the fields, which
corresponds to take the trace over the environmental degrees of freedom in open
quantum system theory, the reduced system state is given by $\rho(t)=\Tr_{\rm
env} \left\{U(t)\,\rho(0) \otimes|e(0)\rangle\langle e(0)|
U(t)^\dagger\right\}$, with $\rho(0)$ the initial state of the oscillator.
Thanks to \eref{eq:U} the reduced dynamics of the mechanical oscillator can be
shown to obey exactly the master equation \eref{Lcala} \cite{Bar06}. Further,
we have the important relations
\begin{equation} \label{qmeans}\eqalign{
\langle e(0)| B_{\rm th}(t)B_{\rm th}^\dagger(s)e(0)\rangle=\left(N+1 \right){\min\{t,s\}},
\\ \langle e(0)| B_{\rm th}^\dagger(t)B_{\rm th}(s)e(0)\rangle=N{\min\{t,s\}},\\ \langle e(0)|
B_{\rm th}(t)B_{\rm th}(s)e(0)\rangle=0.}
\end{equation}
It is worth noticing that the thermal parameter $N$ does not appear in the
commutation rules of the field $B_{\rm th}$, but rather in the quantum
correlations \eref{qmeans}. This expresses the fact that $N$ depends on the
``state'' of the field or, more precisely, $N$ determines a non-Fock
representation of the canonical commutation relations. Note furthermore that
the vacuum $e(0)$ is not annihilated by the fields $B_{\rm th}(t)$, but it
plays the role of a thermal state \cite[Sect.\ 6]{AttJ07}; no vacuum state
exists for a non-Fock Bose field.

\subsection{Quantum Langevin equations and input-output relations}\label{sec:qL}
Relying on the previously introduced formalism we are now in the position to
obtain the so-called quantum Langevin equations, providing the stochastic
evolution for the system observables in the Heisenberg picture. For a generic
system operator $X$ we denote as usual the Heisenberg picture as
$X(t)=U(t)^\dagger XU(t)$, with $U(t)$ the unitary operator describing the
closed dynamics of system and environment. Differentiating this expression by
the rules of quantum stochastic calculus, essentially summarized by the It\^o
table \eref{Ito}, one obtains the quantum Langevin equations for the relevant
system operators, namely for the mode operator
\begin{equation}
\rmd a_\rmm (t)=-\left(\rmi \om+\frac \mgam  2\right) a_\rmm (t)\rmd t -
\sqrt{\mgam } \,\rmd B_{\rm th}(t).
\end{equation}
By \eref{qp-a} we get also the equivalent equations for position and momentum
\begin{eqnarray}\label{eq:q}
\rmd q(t)&=\frac {p(t)} m\,\rmd t +\rmd C_q(t),
\\  \label{eq:p}
\rmd p(t)&=-\left(m\Omq  q(t) +\mgam  p(t)\right) \rmd t+ \rmd C_p(t),
\end{eqnarray}
in which we have introduced the Hermitian quantum noises
\begin{eqnarray}\label{C_qp}\eqalign{
 C_q(t)&= -\sqrt{\frac{\hbar\mgam }{2m\om}} \left(\overline\tau\, B_{\rm th}(t)+  \tau B_{\rm th}^\dagger(t)\right),
\cr
C_p(t)&=\rmi\Om  \sqrt{\frac{m\hbar \mgam }{2\om}}\left(B_{\rm th}(t)- B_{\rm th}^\dagger(t)\right),}
\end{eqnarray}
where $\tau$ is the pase factor defined in \eref{atau}. By \eref{eq:3} the new
noises obey the commutation rules
\begin{equation}\label{[CC]}\fl
\left[ C_q(t), C_p(s)\right]=  \rmi\hbar\mgam \, {\min\{t,s\}}, \qquad
\left[ C_q(t), C_q(s)\right]=\left[ C_p(t), C_p(s)\right]=0.
\end{equation}
A fundamental advantage of the considered formalism is that, thanks to the
unitarity of $U(t)$, the transformation $X\mapsto U(t)^\dagger XU(t)$ preserves
all the commutation rules among system observables, in particular the
Heisenberg relations between position and momentum, as can be checked also
directly relying on \eref{[CC]}. Warranting preservation of these fundamental
commutation relations is indeed a basic step in providing a true quantum
description of a dissipative dynamics \cite[Chapts.\ 1, 3]{GarZ00}.

We now consider the Heisenberg picture for the thermal fields and we define
\begin{equation}
  \label{eq:2}
  B_{\rm th}^{\mathrm{out}}(t)=U(t)^\dagger B_{\rm th}(t)U(t).
\end{equation}
While $B_{\rm th}(t)$ represents the field before the interaction with the
oscillator, the so-called \emph{input field}, $B_{\rm th}^{\mathrm{out}}(t)$ is
the field after the interaction, the so-called \emph{output field}. We stress
in particular that an important consequence of the Hudson-Parthasarathy
equation is the identity $B_{\rm th}^{\mathrm{out}}(t)=U(T)^\dagger B_{\rm
th}(t)U(T)$, $\forall T\geq t$, which warrants the fact that the output fields
obey the same commutation relations as the input fields, namely \eref{eq:3}; in
other words, both the input and the output fields behave as free fields. By
differentiating the three contributions in $U(t)^\dagger B_{\rm th}(t)U(t)$
according to the It\^o rules, one gets the \emph{input-output relation}
\begin{equation}\label{in-out}
\rmd B_{\rm th}^{\mathrm{out}}(t)=\rmd B_{\rm th}(t)+\sqrt{\mgam }\,a_\rmm (t)\,\rmd t.
\end{equation}
The linearity of the Heisenberg equations of motion allows for an explicit
solution
\begin{equation}\label{a_rmm(t)}
a_\rmm (t)=\rme^{-\left(\rmi \om +\frac \mgam  2 \right)t}a_\rmm -\sqrt \mgam
\int_0^t\rme^{-\left(\rmi \om +\frac \mgam  2 \right)\left(t-s\right)}\rmd B_{\rm th}(s),
\end{equation}
\begin{eqnarray}\nonumber \fl
B_{\rm th}^{\rm out}(t)=-\frac{\frac \mgam  2-\rmi \om} {\frac \mgam  2+\rmi \om}\,
B_{\rm th}(t)+ \frac \mgam  {\frac \mgam  2 +\rmi \om} \int_0^t\rme^{-\left(\rmi \om
+\frac \mgam  2 \right)\left(t-s\right)} \rmd B_{\rm th}(s) \\ {}+\frac{\sqrt \mgam  }
{\frac \mgam  2+\rmi \om} \left(1-\rme^{-\left(\rmi \om +\frac \mgam  2\right)t}\right) a_\rmm.
\end{eqnarray}
The explicit expressions for $q(t)$ and $p(t)$ can be easily obtained from
\eref{qp-a} and \eref{a_rmm(t)}.

\subsection{Field state and non-Markovian dynamics}\label{sec:n-M}
In the Markovian approximation considered so far, the temperature enters the
theory only through the parameter $N$ defined in \eref{def:N}.  This
approximation can be described stating that the system actually sees a flat
noise spectrum, or more precisely the system is only affected by the value of
the bath spectrum at the frequency $\om$.  A more general and physically more
realistic situation is to allow for a structured noise spectrum and this can be
achieved without any modification of the unitary dynamics \eref{eq:U} and of
the related Langevin equations and input-output relations. To this aim it is
enough to change the state of the field by taking mixtures of coherent states
\cite{BarPer02,Bar06}. Let us note that considering such a mixture of coherent
states for the description of the state of the field is actually analogous to
consider a state with a regular $P$-representation in the case of discrete
modes (see e.g. \cite{GarZ00}), as explained below. This modification is new in
the context of quantum stochastic calculus and will imply that the reduced
dynamics of the oscillator is no more Markovian, in the sense that a closed
master equation in Lindblad form for the statistical operator cannot be
obtained.

\subsubsection{The field state.}
In order to consider a more general field state let us first introduce the Weyl
operators \cite{Parthas92,Bar06} for the Fock $A$-fields, defined as
\begin{equation*}
\Wcal_A(g)= \exp\biggl\{\sum_{k=1}^2\int_0^{+\infty} g_k(s)
\, \rmd A_k^\dagger(s) - \text{h.c.}\biggr\},
\end{equation*}
with $g_k$ square integrable functions. The operator $\Wcal_A(g)$ is unitary
and the property $A_k(t)\Wcal_A(g)e(0)= \int_0^t \rmd s g_k(s) \Wcal_A(g)e(0)$
holds, so that the action of a Weyl operator on the Fock vacuum gives a
coherent state. Therefore  $\Wcal_A(g)$ is nothing but a displacement operator
for the Bose fields \cite{BarG13}. Relying on \eref{thfield}, we can introduce
a Weyl operator also for the $B$-field
\begin{equation}
\Wcal_T(f)=\exp\biggl\{\int_0^T f(s) \, \rmd B_{\rm th}^\dagger(s) - \text{h.c.}\biggr\},
\end{equation}
where  $f$ is a locally square integrable function and $T$ denotes a suitable
large time, which we will let tend to infinity in the final formulae describing
the quantities of direct physical interest. Now, $\Wcal_T(f)e(0)$ is not a
coherent state for the $B_{\rm th}$-field, but its relevant moments can be
computed by using the $A$-field representation \eref{thfield}:
\begin{equation}\label{<B>}\eqalign{
\fl
\langle \Wcal_T(f)  e(0)|B_{\rm th}(t) \Wcal_T(f) e(0)\rangle &=\int_0^t f(r)\,\rmd r,
\cr
\fl
\langle \Wcal_T(f)  e(0)|B_{\rm th}(t)B_{\rm th}(s) \Wcal_T(f) e(0) \rangle&=
\int_0^t f(u)\,\rmd u\int_0^s f(r)\,\rmd r,
\cr
\fl
\langle \Wcal_T(f)e(0)|B_{\rm th}^\dagger(s)B_{\rm th}(t)\, \Wcal_T(f)e(0)\rangle
&=N{\min\{t,s\}}+\int_0^t f(u)\,\rmd u\int_0^s \overline{f(r)}\,\rmd r,
\cr
\fl
\langle \Wcal_T(f)e(0)|B_{\rm th}(t)B_{\rm th}^\dagger(s)\, \Wcal_T(f)e(0)\rangle
&=\left(N+1\right){\min\{t,s\}}+\int_0^t f(u)\,\rmd u\int_0^s \overline{f(r)}\,\rmd r.}
\end{equation}

A crucial step is now to consider $f$ to be a random process and to take the
state of the field characterizing the environment to be
\begin{equation}\label{newS}
\envS  =\Ebb\left[\Wcal_T(f)|e(0)\rangle\langle e(0)|\Wcal_T(f)^\dagger\right].
\end{equation}
Again, in the final formulae we will take the limit $T\to+\infty$. This is
nothing but an analogue of the regular $P$-representation for the case of
discrete modes. Indeed, in the case of a single mode the Glauber-Sudarshan
$P$-representation of a state $\rho$ \cite[Section 4.4.3]{GarZ00} is defined by
$\rho=\int \rmd^2\alpha\, P(\alpha,\alpha^*)|\alpha\rangle\langle \alpha|$. If
the pseudo-density $P$ is allowed to become negative and singular, then any
state can be represented in this form. When $P$ is a true probability density,
one speaks of a regular $P$-representation and the Glauber-Sudarshan formula
describes mixtures of coherent states, including in particular thermal states
\cite[p.\ 113]{GarZ00}. In a probabilistic language, which is more suitable for
generalizations to stochastic processes and fields, the fact that a state
$\rho$ has a regular $P$-representation can be rephrased by saying that it can
be written as the expectation value $\rho=\Ebb[|\alpha\rangle\langle \alpha|]$,
with $\alpha$ a complex random variable. In order to construct a thermal state
it is enough to consider the case in which the distribution of $\alpha$ is
Gaussian with vanishing mean $ \Ebb[\alpha]=0$ and second moments
$\Ebb[\alpha^2]=0$, $ \Ebb[\abs\alpha^2]=\sigma^2$. Then,
$\rho=\Ebb[|\alpha\rangle\langle \alpha|]$ turns out to be a thermal state
\cite[Section 4.4.5]{GarZ00}.

By analogy, to construct a thermal field state with a general thermal spectrum
we take $f$ to be a Gaussian stationary stochastic process with vanishing mean,
$\Ebb[f(t)]=0$, and correlation functions
\begin{equation}
\Ebb[f(t)\,f(s)]=0, \qquad \Ebb[\overline{f(t)}\,f(s)]=:G(t-s).
\end{equation}
Thanks to stationarity, the function $G(t)$ is positive definite, so that
according to Bochner's theorem \cite{ReedSimonII} its Fourier transform
\begin{equation}
  \hat G(\nu)=\int_{-\infty}^{+\infty}\rme^{-\rmi\nu t}G(t)\,\rmd t
\end{equation}
is a positive function, which we assume to be absolutely integrable, thus
implying a finite power spectral density for the process. Since the field state
$\envS $, defined by \eref{newS}, turns out to be Gaussian, we can characterize
it through the means and the correlations of the thermal field $B_{\rm th}$,
which are immediately obtained from \eref{<B>} and the properties of the
process $f$. The only non zero contributions are given by
\begin{equation}\label{BBcorr}\eqalign{
\fl
  \langle B_{\rm th}^\dagger(s)B_{\rm th}(t)\ranenv
  =N{\min\{t,s\}}+\int_0^{t} \rmd
  u\int_0^{s}\rmd r\,G(r-u),
\\
\fl
\langle B_{\rm th}(t)B_{\rm th}^\dagger(s)\ranenv=\left(N+1\right){\min\{t,s\}}+
\int_0^{t} \rmd u\int_0^{s}\rmd r\,G(r-u).}
\end{equation}

To better grasp the physical content of the new state and of the formulae
\eref{BBcorr} let us introduce a set of field modes as in \cite{BarG13}. Using
a complete orthonormal set $\{h_n\}$ in $L^2(\Rbb)$, we can expand the field in
terms of discrete independent temporal modes by defining them as
\[
c_{h_n}=\int_{-\infty}^{+\infty}\overline{h_n(t)}\,\rmd B_{\rm th}(t).
\]
We then obtain $\langle   c_{h_n} \ranenv= 0$, $\langle c_{h_n}^{\,2}
\ranenv=0$, together with
\[
\lim_{T\to +\infty}\langle c_{h_n}^\dagger c_{h_n}\ranenv=N+\Ebb\left[
\abs{\langle {h_n}|f\rangle_{L^2}}^2\right]=\int_{-\infty}^{+\infty}
\abs{\hat h_n(\nu)}^2 N(\nu)\,\rmd \nu, \]
where we have defined the positive quantity
\begin{equation}\label{N(nu)}
       N(\nu)=N+\hat G(\nu)
\end{equation}
and  ${\hat h_n(\nu)}$ is the Fourier transform of $h_n(t)$; by normalization
$\int_{-\infty}^{+\infty}|{\hat h_n(\nu)}|^2\rmd \nu=1$. So, the reduced state
of the single mode $c_{h_n}$ is exactly a thermal state expressed in the
$P$-representation. If we take $h_1$ and $h_2$ having non overlapping Fourier
transforms we also get $\lim_{T\to +\infty}\langle c_{h_1}^\dagger
c_{h_2}\ranenv=0$, which means that these two modes are independent. Then,
\emph{$N(\nu)$ is naturally interpreted as the mean number of  phonons in a
given field mode $c_h$ well peaked around the value $\nu$} of the frequency and
field modes with different frequencies are independent. A value of $N(\nu)$
different from zero in a neighbourhood of $\nu$ implies that the mechanical
oscillator can absorb from the bath phonons with energy around $\hbar \nu$. On
the contrary, the approximations are such that the oscillator can emit phonons
of any frequency, even when $N(\nu)=0$. The physically relevant quantity is now
the combination of the two non negative contributions $N$ and $\hat G(\nu)$,
rather than the values of the individual quantities. Note that the Markovian
reduced dynamics of \Sref{sec:MELin} can be obtained either by considering the
non-Fock representation for the thermal field, thus assuming a strictly
positive $N>0$ in \eref{thfield} and taking $\hat G(\nu)\equiv 0$, or
equivalently by considering a standard Fock representation and formally taking
the limit of constant spectrum $\hat G(\nu)$ in all the physical quantities.

\subsubsection{The equilibrium state of the mechanical oscillator.}
According to the definition of reduced dynamics, the time evolved state of the
mechanical oscillator is still obtained by taking the partial trace with
respect to the field degrees of freedom $\rho(t)=\lim_{T\to +\infty}\Tr_{\rm
env} \left\{U(t) \left(\rho(0)\otimes \envS
  \right)U(t)^\dagger\right\}$.  However, at variance with the case in
which the state of the field was taken to be the $A$-field vacuum, by taking
the time derivative of this expression no closed evolution equation is obtained
unless $N(\nu)$ is constant. Not to have a closed time-homogeneous equation for
the reduced dynamics is indeed a signature of the non-Markov features of such a
dynamics.

In spite of the difficulty of not having a closed master equation, the study of
the reduced equilibrium state, namely $\rho_{\rm eq}= \lim_{t\to +\infty
}\rho(t)$, can still be afforded and its expression enlightens the physical
role of the various parameters. Indeed, thanks to the requirement
$\Ebb[f(t)\,f(s)]=0$, one has that equipartition in the sense of \eref{equip}
still holds. Starting from the explicit forms of position and momentum in the
Heisenberg picture (see \eref{a_rmm(t)}, \eref{qp-a}) one can check that the
equilibrium mean values of position and momentum remain equal to zero, while
the variances are still of the form \eref{eqvalues} with $N$ replaced by the
effective mean number of excitations
\begin{equation}\label{eq:7}
N_{\rm eff}=\frac \mgam  {2\pi} \int_{-\infty}^{+\infty}\frac{N(\nu)}{\frac{\mgamq }4
+\left(\nu-\om\right)^2}\,\rmd \nu.
\end{equation}
Notice that if the quantity $N(\nu)$ introduced in \eref{N(nu)} is taken to be
the constant $N$, corresponding to the Markovian case, then $N_{\rm eff}=N$.
This result suggests that the final Markov approximation should be valid when
$\hat G(\nu)$ is approximately constant in a neighbourhood of $\om$.  In fact
equation \eref{eq:7} represents a smearing of $N(\nu)$ around the frequency of
the mechanical oscillator $\om$, the more peaked the smaller the damping
constant $\mgam $. Non-Markovian effects can only be relevant if $\hat G(\nu)$
appreciably varies in a neighbourhood of width $\mgam $ around $\om$, being
suppressed with decreasing $\mgam $.

Since the equilibrium state is necessarily Gaussian, by comparing \eref{eq:7}
with \eref{eqstate} we get that the new equilibrium state is again a Gibbs
state with respect to the same Hamiltonian  $H_\rmm$, but with an effective
inverse temperature $\beta_{\rm eff}$ defined by setting $N_{\rm eff} =
({\rme^{\beta_{\rm eff}\hbar\om}-1})^{-1}$.

\subsection{Properties of the quantum noises and quantum stochastic Newton equation}
\label{sec:Newt}

Let us now come back to the quantum Langevin equations for the position and
momentum operators, so as to better understand their physical meaning and the
role of the noises. In order to study the properties of the noises we transform
the Langevin equations \eref{eq:q}, \eref{eq:p} in the form of a stochastic
Newton equation.

To this aim we first have to consider the quantum noises \eref{C_qp} appearing
in these quantum Langevin equations. The commutation relations \eref{[CC]} for
these noises, which are state independent, guarantee the preservation of the
canonical Heisenberg commutation relations. Their quantum correlations do
instead reflect the physical properties of the field state $\envS$ and can be
obtained starting from the $B$-correlations \eref{BBcorr}. Note that Langevin
equations for a mechanical oscillator of the same form and with two noises
obeying the same commutations rules \eref{[CC]} were used also in
\cite{JacTWS99}; however, the two point correlations used in \cite{JacTWS99},
taken from \cite{Dio93a}, arise from approximations in the Caldeira-Legget
model which are valid only for medium/high temperatures, while in the present
treatment they are deduced from the state of the phonon environment and are
valid at any temperature.

We stress the fact that in the present formulation the momentum operator is not
related to the time derivative of the position operator according to the
classical relation, but rather through \eref{eq:q} where the quantum noise
$C_q(t)$ explicitly appears. However, the connection to the classical
formulation is not completely lost. In fact from \eref{eq:q} we can derive the
relation
\[
\frac{q(t_2)-q(t_1)}{t_2-t_1} - \frac 1 {t_2-t_1}\int_{t_1}^{t_2} \frac {p(t)}m \,\rmd t
= \frac{C_q(t_2)-C_q(t_1)}{t_2-t_1}.
\]
By \eref{C_qp} and \eref{BBcorr}, the mean value of the r.h.s.\ of the equation
above vanishes, while its variance is given by
\[ \fl
\frac{\langle \left(C_q(t_2)-C_q(t_1)\right)^2\ranenv }{\left(t_2-t_1\right)^2}=
\frac{\hbar \mgam  }{m \om \left(t_2-t_1\right)}\biggl(\frac 12+\int_{-\infty}^{+\infty}
\frac {2\left(\sin \frac{\nu\left(t_2-t_1\right)}2\right)^2}{\pi \nu^2\left(t_2-t_1\right)}
\, N(\nu)\,\rmd \nu\biggr),
\]
so that in particular also the variance goes to zero for growing $t_2-t_1$.
Then the quantity $v(t)= p(t)/m$ can actually be interpreted as the ``coarse
grained'' velocity of the mechanical oscillator.

If we use the formal field densities $b_{\rm th}(t)$, $b_{\rm th}^\dagger(t)$,
with commutation rules \eref{deltacomm}, take as starting point the quantum
Langevin equations \eref{eq:q} and \eref{eq:p} and eliminate the momentum, we
can rewrite the quantum Langevin equations in the Newton form:
\begin{equation}\label{qddot}
m\ddot q(t)+m\mgam  \dot q(t) +m\Omq q(t)=\xi(t),
\end{equation}
where we have introduced the formally Hermitian quantum noise $\xi(t)$
\begin{equation}
\xi(t)= \dot C_p(t)+m\mgam  \dot C_q(t) +m \ddot C_q(t). \label{noisexi}
\end{equation}
Most importantly the commutation relations for this noise take the singular
expression
\begin{equation}
\label{commXi}
[\xi(t),\xi(s)]=2\rmi m\hbar \mgam  \,\frac {\partial\ }{\partial t} \,\delta(t-s).
\end{equation}
While the expectation value of this noise with respect to the field state
$\envS$ is zero, its symmetrized correlation function can be computed from the
relations
\begin{eqnarray}\fl\nonumber
\frac{m}{\hbar\mgam }\frac{\partial^2 \ }{\partial t\partial s}\langle
C_q(t)C_q(s)  \ranenv = \frac1{m\hbar\mgam  \Omq}\frac{\partial^2 \
}{\partial t\partial s}\langle   C_p(t)C_p(s)  \ranenv   \\
{}=\frac{1}{\om}\left\{\left(N+\frac 1 2 \right)\delta(t-s) +\RE G(t-s)\right\},\label{qqpp}
\end{eqnarray}
\begin{eqnarray}\fl\nonumber
\frac {1}{\hbar\mgamq}\,\frac{\partial^2 \ }{\partial t\partial
  s}\langle \left\{  C_q(t),C_p(s)\right\}  \ranenv
=-\frac{1}{\om}\left\{\left(N+\frac 1 2 \right)\delta(t-s) +\RE G(t-s)\right\}\\ {}+\frac{2}\mgam
 \,\IM G(t-s)\label{+qp}
\end{eqnarray}
and has the expression
\begin{eqnarray}\nonumber\fl
\frac 1 2 \langle \{\xi(t),\xi(s)\}\ranenv = \frac{m\hbar  \mgam }{\om}\left( \Omq
+\frac{\partial^2 \ }{\partial t \partial s}\right) \left[\left(N+\frac 1 2 \right)
\delta(t-s)+\RE G(t-s)\right]\\ {}+2m\hbar  \mgam \,\frac{\partial \ }{\partial t}
\IM G(t-s).\label{+xi}
\end{eqnarray}
Note that \eref{qddot} and \eref{commXi} were already introduced in
\cite[Sect.\ 3.1.2]{GarZ00} and \cite{GioV01}, where the commutation rules
\eref{commXi} were actually enforced by the requirement of preservation of the
commutation rules between position and momentum. However, at variance with
previous approaches, here we have provided an explicit construction of the
quantum noise $\xi(t)$ in terms of a quantum Bose field, based on a rigorous
mathematical construction.

We stress the fact that the stochastic Newton equation \eref{qddot} is
mathematically purely formal due to the presence in \eref{noisexi} of $\ddot
C_q(t)$, which contains the formal derivative $\dot b_{\rm th}(t)$ and its
adjoint.  Moreover, if one were to take \eref{qddot}, \eref{commXi} and
\eref{+xi} as starting point for the construction of the quantum Langevin
equations for position and momentum, then one should complete \eref{qddot},
which is an equation for $q(t)$ only, with a suitable definition of $p(t)$. The
standard choice in this respect, considered for instance in
\cite{GioV01,PGKBBA06,Genes07}, is to take $p(t)=m \dot q(t)$. This works out
fine as far as the commutation relations of position and momentum are
concerned. However, in this case the equation of motion \eref{qddot} and the
structure of the noise $\xi(t)$ obeying \eref{noisexi} imply that $\dot q(t)$
contains singular quantum fluctuations, so that it is not a well defined
operator. Also $p$ is then not a well defined operator and its variance is
actually infinite. Then, one has to regularize the momentum, by subtracting the
noise responsible of this divergence; this is what our construction  does. The
identification of the momentum is given implicitly through the first canonical
equation \eref{eq:q}, which corresponds to the coarse grained velocity, as
discussed above. No divergency appears because the whole construction is based
on the well defined unitary evolution \eref{eq:U}.

\subsubsection{Consistency of the quantum noises.}\label{sec:consistency}
It is important to stress that if a set of quantum Langevin equations is
considered as starting point for the description of a stochastic quantum
dynamics, commutations rules and symmetrized correlations of the noises cannot
be given arbitrarily. In particular, independently of the considered system, if
$\{\xi_i(t)\}$ is a set of operator valued noises, the quantum correlation
function $\langle \xi_i(t)^\dagger \xi_j(t')\ranenv  $ has to be \emph{positive
definite} \cite{ReedSimonII}, in the sense that
\begin{equation}\label{defpos}
\sum_{ij}\int_0^{+\infty}\rmd t\int_0^{+\infty}\rmd t'\,\overline{h_i(t)}\langle
\xi_i(t)^\dagger \xi_j(t')\ranenv  h_j(t')\geq 0,
\end{equation}
for every choice of the ``smooth'' test functions $\{ h_i \}$. Since we can
always write
\begin{equation}
  \xi_i(t)^\dagger \xi_j(t')=\frac 1 2 \left\{\xi_i(t)^\dagger ,\,\xi_j(t')\right\}
  +\frac 1 2 \left[\xi_i(t)^\dagger ,\,\xi_j(t')\right],
\end{equation}
the necessary positivity condition introduced above becomes a consistency
condition between commutation rules and symmetrized correlations.

Relying on \eref{qqpp}, \eref{+qp}, as well as the commutation relations
\eref{[CC]} for the noises $C_q$ and $C_p$, one can immediately check this fact
for the model at hand. Also for the singular noise $\xi$ constrained by
\eref{noisexi} one can show that the expression $\langle \xi(t)\xi(s)\ranenv $
is positive definite. These results are due to the fact that the noise fields
have here been explicitly constructed in terms of the quantum Bose fields, so
that commutation rules and correlations are not postulated, but rather follow
from the mathematical expression of the model.

\subsubsection{The noise correlations.}
For the model at hand we denote the Fourier transform of the correlation of the
noise $\xi$ by
\begin{equation}\label{def:Rnu}
\hat R(\nu)=\frac 1 2\int_{-\infty}^{+\infty}\rmd t\,\rme^{-\rmi\nu t}  \langle \{\xi(t+s),\xi(s)\}\ranenv,
\end{equation}
so that according to \eref{N(nu)} and \eref{+xi} it reads
\begin{equation}
\hat R(\nu)=\frac{m \hbar \mgam } {2\om}\left(\frac{\mgamq }4+ \left(\om +\nu\right)^2\right)
\left(N(\nu)+\frac 1 2 \right) +(\nu \to -\nu), \label{1.corr.xi}
\end{equation}
where $(\nu \to -\nu)$ means to add the same contribution with $\nu$ replaced
by $-\nu$. Note in particular that $\hat R(\nu)$ is an even function of the
frequency.

In our treatment, which gives rise to the expression \eref{1.corr.xi}, the
interaction with the environment is described in terms of exchange of quanta
with the bosonic field representing the phonons, see \eref{TU(t)}. In models in
which the system of interest is coupled to other harmonic oscillators, by some
approximations it is possible to arrive to a quantum stochastic Newton equation
like \eref{qddot}, but with a different noise spectrum. A reference expression
often considered in the literature \cite[(3.3.9)]{GarZ00}, \cite{GioV01} for
the quantity $\hat R$ is given by
\begin{equation}\label{GZnu}
\hat R_{GZ}(\nu)=\pi \hbar J(\nu) \coth
\frac{\beta\hbar \nu}{2}, \qquad J(\nu):=\frac m \pi \, k(\nu) \nu.
\end{equation}
The quantity $k(\nu)$ contains information on both coupling constant and
density of modes of the bath in a neighbourhood of the frequency $\nu$;
$J(\nu)$ is often called \emph{spectral density}. Also in the case of this
choice of the noise correlations, it is possible to show that the equilibrium
mean of $\dot q(t)^2$ diverges and therefore the identification of the momentum
with $m\dot q(t)$ is not possible, but some regularization is needed. A typical
choice in this context is $k(\nu)=\mgam $ \cite[(3.1.1)]{GarZ00},
\cite{GioV01,PGKBBA06,Genes07,GMVT09}; this is equivalent to $J(\nu)\propto
\nu$, which is known as \emph{Ohmic spectral density}. The function $k(\nu)$
must be even due to the definition \eref{def:Rnu} and stationarity. Also the
commutations rules \eref{commXi} and the positivity requirement \eref{defpos}
still have to hold, leading to the requirement $\hat R_{GZ}(\nu)-m \hbar\mgam
\abs\nu \geq 0$, satisfied at any temperature by taking $k(\nu)\geq\mgam >0$
for all $\nu$. This requirement tells us that in order to have a consistent
model satisfying \eref{qddot}, at least at large times, and preserving
Heisenberg commutation relations one cannot consider a spectral density with
gaps inside the expression \eref{GZnu} of $\hat R_{GZ}$.

By a suitable choice of $N(\nu)$, it is possible to get $\hat R(\nu)$  very
similar to $\hat R_{GZ}(\nu)$, apart from the low temperature limit. For
instance, by taking
\begin{equation}
N(\nu)=\abs\nu k(\nu)\,\frac{2\om}{\left(\Omq+\nu^2\right)\mgam
\left(\rme^{\beta \hbar \abs \nu}-1\right)}, \label{N(nu)even}
\end{equation}
we obtain
\begin{equation}\label{RtoRGZ}
\hat R(\nu)=
\hat R_{GZ}(\nu)+ m\hbar\left[\frac{\mgamq }{2\om}\left(\Omq+\nu^2\right)-k(\nu)\abs\nu\right];
\end{equation}
note that the difference is independent from the temperature. In \eref{RtoRGZ}
the compatibility with the commutation relations is guaranteed by the first
term in the square brackets; so, in \eref{N(nu)even} there is no restriction on
the choice of $k(\nu)$, apart from $k(\nu)\geq 0$, and even spectral gaps can
be introduced.

Besides the case \eref{N(nu)even}, the freedom in the choice of $N(\nu)$ in
\eref{1.corr.xi} allows to model quite different environments, for instance
with a sub-Ohmic or super-Ohmic spectral density \cite{XZWW10,ZLXTN12,YCXZS13},
or with a structured occupation spectrum. In optomechanical systems the
quantity $\hat R(\nu)$ enters experimental quantities, as in the cases of
homodyne/heterodyne detection discussed in Section \ref{sec:optical-spectra};
so, in principle it is possible to test the form of $\hat R(\nu)$. However,
essentially a frequency window of width $\mgam$ around $\om$ is experimentally
relevant. Some results at room temperature \cite{GroTru15} seem to indicate a
non-Ohmic spectral density around $\om$, a very interesting possibility, but
not enough to discriminate between $\hat R_{GZ}(\nu) $ and the form
\eref{RtoRGZ} for $\hat R(\nu)$. At zero temperature, corresponding to
$N(\nu)=0$ in our case \eref{1.corr.xi} and to $\beta \to +\infty$ in
\eref{GZnu}, the difference is most evident, namely
\begin{equation}\label{T=0,R}
\hat R(\nu)= m\hbar \mgam \frac{\Omq +\nu^2}{2\om}
\end{equation}
versus
\begin{equation}\label{T=0,GZ}
\hat R_{GZ}(\nu)=m\hbar k(\nu)\abs\nu, \qquad k(\nu)\geq\mgam.
\end{equation}
These expressions are very different, but to discriminate between them one has
to consider very low temperatures and a ratio $\mgam/\om$ that is not too
small.

\section{Cooling and emission spectra of an optomechanical system}\label{sec:model1}
As an application of the quantum description of a mechanical oscillator
developed so far we consider the simplest optomechanical system
\cite{GioV01,GMVT09,PGKBBA06,Genes07,SN-P13,Chen13}, namely the mechanical
oscillator is a mirror mounted on a cantilever and coupled to the light in an
optical cavity by radiation pressure. The cavity is of high quality, without
thermal dissipation other than the one due to the coupling between cantilever
and phonons and tuned in such a way that only one electromagnetic mode is
relevant. Strong laser light is injected and some light is allowed to leave the
cavity so that
  its spectrum can be analysed.

\subsection{The optomechanical model}
\label{sec:optomechanical-model} The micro-mechanical oscillator (the mirror)
is described by the operators $q$, $p$ as in \eref{qp-a} and by the Hamiltonian
$H_\rmm$ \eref{H0H00}. The cavity mode is described by the operators $a_{\rm
c}$, $a_{\rm c}^\dagger$ and by the free Hamiltonian $\hbar \oc a_{\rm
c}^\dagger a_{\rm c}$. The free electromagnetic field is in a coherent state
describing a perfectly monochromatic laser of frequency $\omega_0$; however we
use the equivalent description of inserting a source term for the cavity mode
in the Hamiltonian and of taking the external field in the vacuum. The final
optomechanical Hamiltonian takes the form
\begin{equation}
H_{\rm om}(t)=H_\rmm+\hbar \oc a_{\rm c}^\dagger a_{\rm c}
-\hbar g_0 q a_{\rm c}^\dagger a_{\rm c}
+\rmi \hbar E\left(a_{\rm c}^\dagger \rme^{-\rmi \omega_0 t}
-a_{\rm c} \rme^{\rmi \omega_0 t}\right).
\end{equation}
Note the trilinear term giving the interaction between the position of the
mirror and the number operator of the photons in the cavity, which represents
the radiation pressure interaction; the coupling constant is usually expressed
as $g_0=\oc/L$, where $L$ is the length of the cavity. The laser power is
$P=\hbar\omega_0 E^2/\cgam  $, where $\cgam$ is the cavity decay rate and $E$
the laser amplitude.

In order to include the cavity mode interacting through radiation pressure with
the mechanical oscillator, as well as the emission and absorption of the light
from the free electromagnetic field, the Hudson-Parthasarathy equation
\eref{eq:U} is modified as follows:
\begin{eqnarray}\fl\nonumber
\rmd U(t)=\biggl\{-\frac\rmi \hbar\,H_{\rm om}(t)\rmd t +
\left(\sqrt{\mgam }\,a_\rmm\,\rmd B_{\rm th}^{\;\dagger}(t) +\sqrt{\cgam }\,a_{\rm c}\,
\rmd B_{\rm em}^{\;\dagger}(t)-\text{h.c.}\right) \\
{}-\frac \mgam {2 }\left(\left(2N+1\right) a_\rmm^\dagger
  a_\rmm+N\right)\rmd t - \frac\cgam  2\, a_{\rm c}^\dagger a_{\rm c} \rmd t \biggr\}U(t).
\label{eq:Umc}
\end{eqnarray}
Here $B_{\rm th}$ is the thermal field satisfying \eref{eq:3}, while $B_{\rm
em}$ is an independent Bose field in the Fock representation, describing the
electromagnetic field outside the cavity. The relevant It\^o rule is $\rmd
B_{\rm em}(t)\rmd B_{\rm
  em}^{\;\dagger}(t)=\rmd t$, while all the other possible products vanish. Now
$U(t)$ is the unitary operator describing the dynamics of the two interacting
oscillators and the fields. The latter are in a factorized state given by the
tensor product of the thermal environment state \eref{newS} and the
electromagnetic vacuum:
\begin{equation}\label{tiledeenvS}
\tilde\envS=\envS\otimes |e_{\rm em}(0)\rangle \langle e_{\rm em}(0)|.
\end{equation}

It is convenient to eliminate the laser frequency working in the rotating frame
and introducing the unitary operator $V(t)=\rme^{\rmi \omega_0 a_{\rm
c}^\dagger a_{\rm c} t}U(t)$, which upon differentiation obeys an equation of
the form \eref{eq:Umc} albeit with $H_{\rm om}(t)$ substituted by
\begin{equation}
H_\rmm+\hbar \Delta_0 a_{\rm c}^\dagger a_{\rm c}-\hbar g_0 q a_{\rm
  c}^\dagger a_{\rm c}+\rmi \hbar E\left(a_{\rm c}^\dagger -a_{\rm c}
\right),
\end{equation}
with $\Delta_0= \oc -\omega_0$ the nominal detuning. For a generic system
operator $X$ we define $X(t)= V(t)^\dagger XV(t)$, so that by differentiating
according to the rules of quantum stochastic calculus, as done in
\Sref{sec:qL}, we get the following quantum Langevin equations
\begin{equation}\fl
\rmd a_{\rm c}(t)
=\biggl(-\left(\rmi \Delta_0+\frac\cgam  2\right) a_{\rm c}(t)+ \rmi g_0q(t)a_{\rm c}(t)
-\rmi E\biggr)\rmd t- \sqrt{\cgam } \,\rme^{\rmi \omega_0 t}\rmd B_{\rm em}(t),
\end{equation}
as well as
\begin{equation}\eqalign{
\rmd q(t)=\frac {p(t)} m\,\rmd t +\rmd C_q(t),
\cr
\rmd p(t)=\left(-m\Omq  q(t) -\mgam  p(t)+\hbar g_0a_{\rm c}^\dagger(t)
a_{\rm c}(t)\right) \rmd t+ \rmd C_p(t),}
\end{equation}
where $C_q$ and $C_p$ are given by \eref{C_qp}. Defining the output fields as
in \eref{eq:2} of \Sref{sec:qL} we have besides \eref{in-out} the input-output
relation for the electromagnetic field
\begin{equation}
\rmd B_{\rm em}^{\rm out}(t)= \rmd B_{\rm em}(t)+\sqrt \cgam  \,\rme^{-\rmi
\omega_0 t}a_{\rm c}(t)\rmd t.
\end{equation}

In the case of a very intense laser, that is $E^2$ large, the dynamics can be
linearized in a neighbourhood of the equilibrium mean values, determined by
self-consistency from the means of the linearized form of the quantum Langevin
equations. The equilibrium mean value of the momentum is zero, while setting $
\zeta=\langle a_{\rm c}(t)\rangle_{\rm eq}$, we find
\begin{equation}\label{eqpos}
\zeta= -\frac{\rmi E}{\frac\cgam  2+\rmi \Delta},\qquad  \langle q\rangle_{\rm eq}
=\frac{\hbar g_0\abs\zeta^2}{m\Omq},
\end{equation}
where we have introduced the effective detuning $\Delta$,
\begin{equation}\label{def:Delta}
\Delta= \Delta_0- g_0 \langle q\rangle_{\rm eq}
=\omega_{\rm c}- g_0 \langle q\rangle_{\rm eq}-\omega_0.
\end{equation}
By inserting the equations \eref{eqpos} into \eref{def:Delta} we obtain the
self-consistency condition
\begin{equation}
m\Omq\left(\Delta-\Delta_0\right)\left(\frac{\cgamq }4 +\Delta^{2}\right)+\hbar g_0^{\,2}E^2=0;
\end{equation}
this cubic equation determines $\Delta$ as a function of the laser parameters
$\Delta_0$ and $E$.

In writing and solving the linearized quantum Langevin equations it is useful
to have dimensionless and selfadjoint system operators. It is therefore
convenient to set
\begin{equation}
\hat q(t)= \sqrt{\frac{m\Om}\hbar}\bigl(q(t)-\langle q\rangle_{\rm eq}\bigr),
\qquad \hat p(t)=\frac{p(t)}{\sqrt{m\hbar \Om}},
\end{equation}
\begin{equation}\fl
X(t)= \frac{\zeta a_{\rm c}^\dagger(t)+\overline{\zeta}\,a_{\rm c}(t)}
{\sqrt 2\abs \zeta}- \sqrt 2\,\abs \zeta,
\qquad
Y(t)= \frac{\rmi\left(\zeta a_{\rm c}^\dagger(t)-\overline{\zeta}\,
a_{\rm c}(t)\right)}{\sqrt 2\abs \zeta}.
\end{equation}
Then, the linearized quantum Langevin equations turn out to be
\begin{equation}\label{eq:x}
\rmd \vec w(t)= A\vec w(t) \rmd t - \rmd \vec Q(t),
\qquad \vec w(t)= \left( \hat q(t),\, \hat p(t),\, X(t),\,  Y(t)\right)^{\rm T},
\end{equation}
where the superscript T means transposition and the dynamical matrix is given
by
\begin{equation}  \label{dynA}
A=\pmatrix{ A_\rmm & A_{\rm mc} \cr A_{\rm mc} &A_{\rm c} }, \qquad
A_{\rm mc}=\pmatrix{ 0& 0 \cr G\sqrt{\om/\Om} & 0 },
\end{equation}
\begin{equation}\label{dynAA}
A_\rmm=\pmatrix{ 0 & \Om \cr -\Om & -\mgam},  \qquad A_{\rm c}=
\pmatrix{ -\cgam/2& \Delta\cr -\Delta & -\cgam/2 }.
\end{equation}
The quantity $G$, having the dimension of a frequency, will play the role of
effective coupling constant and is given by
\begin{equation}\label{GG}
G=g_0\abs{\zeta}\sqrt{\frac{2\hbar}{m\om}}\,,
\end{equation}
so that in particular it depends on the effective detuning $\Delta$ through
$\zeta$ given in \eref{eqpos}.  The vector of noises is given by the following
field combinations:
\begin{equation}\fl \phantom{vv} \quad
Q_1(t)=\tau\sqrt{\frac{\mgam\Om}{ 2\om}}\, B_{\rm th}^{\;\dagger} (t)
+\text{h.c.}, \qquad Q_2(t)=\rmi\sqrt{\frac{\mgam\Om}{ 2\om}}\,
 B_{\rm th}^{\;\dagger} (t)+\text{h.c.},
\end{equation}
\begin{equation}\eqalign{
Q_3(t)=\rme^{\rmi \arg \zeta}\sqrt{\frac\cgam  2}\int_0^t\rme^{-\rmi \omega_0 s}
\rmd B_{\rm em}^{\;\dagger} (s)+\text{h.c.},
\cr
Q_4(t)= \rmi\rme^{\rmi \arg \zeta}\sqrt{\frac\cgam  2}\int_0^t\rme^{-\rmi \omega_0 s}
\rmd B_{\rm em}^{\;\dagger} (s)+\text{h.c.},}
\end{equation}
where $\tau$ is the phase factor defined in \eref{atau} and the quadratures
$Q_1(t)$ and $Q_2(t)$, apart from a multiplicative factor due to the change of
dimensions, coincide with the noises \eref{C_qp}.

Note the different structure of the two dynamical sub-matrices in \eref{dynAA}.
Indeed the former describes a mechanical oscillator and the latter an optical
mode, corresponding to different interactions as discussed in \Sref{sec:MELin}.
The same choice is taken, for instance, in
\cite{JacTWS99,GioV01,PGKBBA06,Genes07,GMVT09,Pontin14}, but not in
\cite{SN-P13,asp14,JJ14}.

The linearization around the equilibrium state is meaningful provided one can
ensure the existence of such a state. Its stability conditions can be obtained
by applying the Routh-Hurwitz criterion to the equations for the mean values,
which correspond to the system \eref{eq:x} with the noise term $\rmd \vec Q(t)$
suppressed; the result is the couple of conditions
\begin{equation}
G^2\om\Delta < \Omq\left(\frac{\cgamq }4
+\Delta^{2}\right), \label{HRcond1}
\end{equation}
for $\Delta>0$, and
\begin{equation}\fl
G^2\om\abs\Delta< \frac{\cgam\mgam}{\cgam+\mgam}\left[
\cgam \Omq+\mgam \left(\frac{\cgamq }4+\Delta^{2}\right)+\frac{\left(\Omq-\frac{\cgamq }4
-\Delta^{2}\right)^2}{\mgam +\cgam }\right],\label{HRcond2}
\end{equation}
for $\Delta<0$; there is no restriction for $\Delta=0$. The same stability
conditions have been found in \cite{Genes07}, as their equations for the mean
values agree with ours.

\subsection{Energy fluctuations and laser cooling}

To introduce the fluctuation spectra of position and momentum of the mechanical
oscillator we use a formulation tailored for (classical or quantum) processes
starting at time zero and we define the \emph{gated Fourier transforms}
\cite{asp14}
\begin{equation}\label{Bnu}
\hat B_i^T(\nu)=\frac 1 {\sqrt T}\int_0^T \rme^{\rmi \nu t}\rmd B_i(t),\qquad i= {\rm th,\ em},
\end{equation}
for the Bose fields as well as for the relevant system operators
\begin{equation}
F_i(T;\nu)=\frac 1 {\sqrt T}\int_0^T\rme^{\rmi \nu t}w_i(t)\rmd t, \quad i=1,2,3,4.
\end{equation}
Here $T$ is a large time going to infinity in the final formulae to recover a
stationary situation. Then, the spectra of the fluctuations of position and
momentum of the mechanical oscillator are defined, in analogy with the
classical case \cite{Howard2002}, by the quantum expectations
\begin{equation}\label{flsp1}\eqalign{
S_q(\nu)= \lim_{T\to +\infty} \frac 12 \langle \left\{F_1(T;\nu),\, F_1(T;-\nu)\right\}\rangle,
\cr
S_p(\nu)= \lim_{T\to +\infty} \frac 12 \langle \left\{F_2(T;\nu),\, F_2(T;-\nu)\right\}\rangle,}
\end{equation}
\begin{equation}\fl
S_{qp}(\nu)= \lim_{T\to +\infty}\frac 14\langle \left\{F_1(T;\nu),\, F_2(T;-\nu)\right\}
+\left\{F_1(T;-\nu),\, F_2(T;\nu)\right\}\rangle.\label{flsp2}
\end{equation}
Let us stress that, while useful, these definitions do not correspond to some
continuous monitoring of position and momentum, even though $S_q(\nu)$ is
directly related to the observed optical spectra as we shall see in
\Sref{sec:optical-spectra}.

The Fourier transformed equations of motion corresponding to \eref{eq:x} can be
solved by purely algebraic manipulations and the vector $\vec F(T;\nu)$ can be
computed; due to the length of the expressions the result is reported in
\ref{app:flu}. To compute the spectra above we need also the field
correlations, which we give in \eref{Enu}.

Due to the  vanishing of the field cross-correlations, the spectra
\eref{flsp1}, \eref{flsp2} decompose in a thermal and a radiation pressure
contribution according to
\begin{equation}
\eqalign{
S_q(\nu)=S_q^{\rm rp}(\nu)+S_q^{\rm th}(\nu),\qquad S_{qp}(\nu)&=S_{qp}^{\rm th}(\nu),
\cr
S_p(\nu)=\frac{\nu^2}\Omq\,
S_q^{\rm rp}(\nu)+S_p^{\rm th}(\nu).}
\end{equation}
By inserting the expressions \eref{sol:F1}, \eref{sol:F2} into the definitions
\eref{flsp1}, \eref{flsp2} and by using \eref{Enu} we get, by some
computations, the expressions for the spectra of the fluctuations:
\begin{equation}\label{Sqrp}
S_q^{\rm rp}(\nu)= \frac{\Om\om G^2\cgam }{2\abs{d(\nu)}^2} \left(\Delta^{2}
+\frac{\cgamq }4 +\nu^2\right),
\end{equation}
\begin{equation}\label{Sqth}
S_q^{\rm th}(\nu)= \frac{\Om\hat R(\nu)}{\hbar m\abs{d(\nu)}^2}\left(\frac{\cgamq }4
+\left(\nu-\Delta\right)^2\right)\left(\frac{\cgamq }4 +\left(\nu+\Delta\right)^2\right),
\end{equation}
\begin{eqnarray}\nonumber\fl
S_p^{\rm th}(\nu)=\frac{\mgam }{2\om\Om\abs {d(\nu)}^2}\biggl\{
\biggl| \left(\Omq+\nu\left(\om-\rmi\,\frac \mgam  2
\right)\right)\left(\Delta^{2}+\left(\frac \cgam  2 -\rmi \nu\right)^2\right)
\\ {}- G^2 \om \Delta\biggr|^2\left(N(\nu)+\frac 1 2 \right)
+(\nu\to -\nu)
\biggr\},
\end{eqnarray}
\begin{eqnarray}\fl
S_{qp}^{\rm th}(\nu)=-\frac \mgam {2\Om}\, S_{q}^{\rm th}(\nu)
+\frac{\mgam G^2 \Delta}{2\abs{d(\nu)}^2}
\biggl\{\left(N(\nu)+\frac 12 \right)
\nonumber
\\
{}\times\left[\frac \mgam  2 \left(\Delta^{2}+\frac {\cgamq }4
-\nu^2\right)-\nu \cgam (\om+\nu)\right]+(\nu \to -\nu)
\biggr\},
\end{eqnarray}
where $(\nu \to -\nu)$ means to add the same contribution with $\nu$ replaced
by $-\nu$ and the quantity $\hat R(\nu)$ is the Fourier transform of the
quantum correlations of the noise given in \eref{1.corr.xi}. The denominator
$d(\nu)$ is the  characteristic polynomial  of the dynamical matrix $A$ given
in \eref{dynA} and \eref{dynAA}:
\begin{eqnarray}\fl
d(\nu)=\det\left(A+\rmi \nu
  \openone\right)=
\left(\left(\nu+\rmi \,\frac {\gamma_{\rm c}}  2 \right)^2-\Delta^{\;2}\right)
\left(\left(\nu+\rmi\,\frac {\gamma_\rmm} 2 \right)^2-\omq\right)- G^2\om\Delta.\label{den}
\end{eqnarray}

Note that the quantities \eref{Sqrp}--\eref{Spth} are non-negative as they
should be in order to give a sensible decomposition of the spectra. Another
useful way to write $S_p^{\rm th}(\nu)$ is by putting in evidence its
difference from $S_q^{\rm th}(\nu)$; the resulting expression is
\begin{eqnarray}\nonumber\fl
S_p^{\rm th}(\nu)- S_q^{\rm th}(\nu) = \frac{\om \mgam
G^2\Delta}{\Om\abs{d(\nu)}^2} \biggl\{\left(N(\nu)+\frac 1 2\right)
\biggl[\frac 1 2\,G^2 \Delta+\nu^2\,\frac{\mgam \cgam }{2\om}
\\
{}+\left(\frac\Omq \om+\nu\right)\left(\nu^2-\Delta^{2}
-\frac{\cgamq }4\right)\biggr] + (\nu \to -\nu)
\biggr\}.\label{Spth}
\end{eqnarray}

\subsubsection{The peaks in the fluctuation spectra.}\label{sec:zeros}

A relevant role in determining the properties of the system is played by the
denominator $d(\nu)$ \eref{den}; indeed, the quantity $\Om[\Delta^2-\left(\nu
+\rmi\cgam /2\right)^2]/d(\nu)$ is sometimes interpreted as the effective
mechanical susceptibility \cite[Eq.\ (17)]{Genes07}. Most importantly note that
the zeros of $d(\nu)$ determine the positions and the widths of the peaks of
the fluctuation spectra: even though in principle they can be obtained by
solving the fourth order algebraic equation $d(\nu)=0$, it is much more
convenient to have simple expressions, even if approximate. An analysis of
these zeros is given in \ref{app:zeros} in the case in which $d(\nu)$ can be
written in the form
\begin{equation}\label{dsolved}
d(\nu)= \left(\left(\nu+\rmi \,\frac {\Gamma_{\rm c}}  2 \right)^2
-\Delta_{\rm eff}^{\;2}\right)\left(\left(\nu+\rmi\,\frac {\Gamma_\rmm} 2 \right)^2
-\omega_{\rm eff}^{\rm m \;2}\right).
\end{equation}
The stability conditions \eref{HRcond1}, \eref{HRcond2} guarantee the strict
positivity of the \emph{effective damping constants} $\Gamma_{\rm c}$ and
$\Gamma_\rmm$. The quantity $\omega_{\rm eff}^{\rm m}$ is known as
\emph{optical spring rigidity}, while the ratio $(\Gamma_\rmm -\mgam )/\mgam $
is called \emph{co-operativity} \cite{SN-P13,asp14}.

An exact expression for the zeros is found when $\Delta=\om$, which allows us
to put into evidence a crossing of the frequencies of the hybridized optical
and mechanical modes \cite{asp14}. If also the condition
\begin{equation}\label{condition2x}
4G^2< (\cgam-\mgam)^2
\end{equation}
holds, the result is
\begin{equation}\label{Gc1}
\Gamma_{\rm c}=\frac{\cgam+\mgam}2+\epsilon\sqrt{2 u^2-2\omq+\frac{(\cgam-\mgam)^2}8},
\end{equation}
\begin{equation}\label{Gm1}
\Gamma_{\rmm}=\frac{\cgam+\mgam}2-\epsilon\sqrt{2 u^2-2\omq+\frac{(\cgam-\mgam)^2}8},
\end{equation}
\begin{equation}\label{Deff1}
\Delta_{\rm eff}^{\;2}=\omega_{\rm eff}^{\rm m \;2}=\frac{\omq + u^2} 2  -\frac {(\cgam-\mgam)^2}{32},
\end{equation}
where
\begin{equation}\fl \label{u+e}
u^2=\sqrt{\left(\omq+\frac{(\cgam-\mgam)^2}{16}\right)^2-G^2 \omq},
\qquad
\epsilon=\cases{+1 & if $\cgam>\mgam$ \cr -1 & if $\cgam<\mgam$}.
\end{equation}

Always for $\Delta=\om$, under the conditions
\begin{equation}\label{condition3x}\fl
\frac{\omq(\cgam-\mgam)^2}4< G^2\omq < \left(\omq +\frac{(\cgam-\mgam)^2}{16}\right)^2,
\qquad
\omq >\frac{(\cgam-\mgam)^2}{16}\,,
\end{equation}
we get instead the result
\begin{equation}\label{G2}
\Gamma_{\rm c}=\Gamma_\rmm=\frac{\cgam+\mgam}2,
\qquad
\Delta_{\rm eff}=\sqrt{x_\pm}, \qquad \omega_{\rm eff}^{\rm m}=\sqrt{x{_\mp}},
\end{equation}
\begin{equation}\label{xpm}
x_\pm= \omq -\frac{(\cgam-\mgam)^2}{16}\pm \om \sqrt{G^2-\frac{(\cgam-\mgam)^2}4}.
\end{equation}
The two alternatives  in \eref{G2} are completely equivalent; there is no
reason to associate the frequency $\sqrt{x_+}$ to the cavity and $\sqrt{x_-}$
to the mechanical oscillator, or viceversa. A striking feature of the case
$\Delta=\om$ is the change in the behaviour of the zeros at the critical point
$\cgam=\bar{\gamma}_{\rm c}$, solution of $G^2=(\cgam-\mgam)^2/4$; recall that
$G^2$ depends on $\cgam$ due to \eref{eqpos} and \eref{GG}.

In the general case, an approximate expression can be obtained under the
conditions
\begin{equation}\label{conditions}
\frac\mgam\cgam \ll 1 ,\qquad \abs{\chi(\Delta)}\ll 1,
\qquad
\abs{\chi(\Delta)}\abs{1-\frac{\Delta^2}\omq-\frac {\cgamq}{4\omq}}\ll 1,
\end{equation}
where
\begin{equation}\label{chi}
\chi(\Delta)=\frac{G^2\om \Delta}{\left(\frac {\left(\cgam-\mgam\right)^2 } 4
+\left(\Delta- \om  \right)^2\right)\left(\frac {\left(\cgam-\mgam\right)^2 } 4
+\left(\Delta+\om  \right)^2\right)}\,.
\end{equation}
The result for the damping constants is
\begin{equation}\label{Gammamc}
\Gamma_{\rmm}\simeq \mgam +\chi(\Delta)\left(\cgam-\mgam\right), \qquad
\Gamma_{\rm c}\simeq \cgam-\chi(\Delta)\left(\cgam-\mgam\right).
\end{equation}
The expressions for $\omega_{\rm eff}^{\rm m \;2}$ and $\Delta_{\rm eff}^{\;2}$
are then obtained by inserting $\Gamma_{\rmm}$ and $\Gamma_{\rm c}$ in the
equations \eref{moduli}. The compatibility conditions \eref{conditions3} have
to hold. As one can see, when $\Delta>0$ (\emph{red detuning}), we have an
increasing of the mechanical damping constant, $\Gamma_\rmm>\mgam$, and a
decreasing of the spring rigidity, $\omega_{\rm eff}^{\rmm}<\om$.

\subsubsection{The mean values at equilibrium.} By integrating in
their frequency dependence the fluctuation spectra one obtains the second
moments of position and momentum in the equilibrium state:
\begin{equation}\label{mean:Sq&p}\fl
\langle q^2\rangle_{\rm eq}-\langle q\rangle_{\rm eq}^2=\frac \hbar {2\pi m \Om}
\int_\Rbb S_q(\nu)\rmd \nu, \qquad \langle p^2\rangle_{\rm eq}=\frac {m\hbar \Om}
{2\pi}\int_\Rbb S_p(\nu)\rmd \nu,
\end{equation}
\begin{equation}\label{mean:Sqp}
\frac 1 2 \langle \left\{q,p\right\}\rangle_{\rm eq}=\frac \hbar {2\pi}\int_\Rbb S_{qp}(\nu)\rmd \nu.
\end{equation}
All these quantities are finite since the integrands behave as $\nu^{-2}$ for
$\abs \nu \to +\infty$. Moreover, the reduced equilibrium state of the
mechanical oscillator is a Gaussian state characterized by \eref{mean:Sq&p} and
\eref{mean:Sqp} together with $\langle q\rangle_{\rm eq}=\hbar
g_0\abs\zeta^2/(m\Omq)$, \ $\langle p\rangle_{\rm eq}=0$.

On the contrary the integral of $\nu^2S_q(\nu)$, which would give the
fluctuations at equilibrium of $\sqrt{{m\Om}/ \hbar}\,\dot q$, does not exist.
This fact is related to the features of the noise in the thermal part and, as
already noticed right before \Sref{sec:consistency}, this noticeably implies
that the standard identification of $m\dot q$ with momentum is not possible.
The expression of $S_q(\nu)$ coincides with the one given in
\cite{PGKBBA06,Genes07}, where however $\hat R_{GZ}(\nu)$ with Ohmic spectral
density appears instead of $\hat R(\nu)$. While in the  case of
\cite{PGKBBA06,Genes07} $S_q(\nu)\asymp \nu^{-3}$, still $\dot q^2$ does not
have a finite mean and also in this case the identification of momentum and
velocity is not possible. Notice that the expressions for $S_p(\nu)$ and
$S_{qp}(\nu)$ have not been obtained before. In particular $S_{qp}(\nu)\neq 0$
implies that the fluctuations of position and momentum are actually correlated.

The mean energy of the harmonic oscillator at equilibrium takes the form
\begin{equation}
\langle H_\rmm\rangle_{\rm eq}=\frac 12\, m \Omq \langle q\rangle_{\rm eq}^2+\langle H\rangle_{\rm fl},
\end{equation}
where the contribution due to fluctuations is given by
\begin{equation}
\langle H\rangle_{\rm fl}=\frac \hbar{4\pi}\int_\Rbb \rmd \nu\left[\Om
\bigl( S_{q}(\nu)+S_{p}(\nu)\bigr) + \mgam  S_{qp}(\nu)\right].
\end{equation}
It is convenient and natural to split this contribution into three distinct
terms, distinguishing a radiation pressure term from the rest and further
dividing the thermal contributions into two, putting into evidence a
contribution which is not proportional to position fluctuations and does not
have a definite sign. We thus introduce the dimensionless quantities
\begin{equation}\label{Nrp}\fl
\mathcal{N}_{\rm rp}=\frac 1{2\pi}\int_\Rbb \frac {\Omq+\nu^2}{2\om\Om}\,S_{q}^{\rm rp}(\nu)\,\rmd \nu,
\qquad
\mathcal{N}_{\rm th}=\frac 1{2\pi}\int_\Rbb   \frac\om\Om\, S_{q}^{\rm
th}(\nu)\,\rmd \nu,
\end{equation}
as well as
\begin{eqnarray}\fl\nonumber
\mathcal{M}_{\rm th}(\Delta)=\frac 1{2\pi}\int_\Rbb \rmd \nu\, \frac{ G^2  \mgam\Delta}{2\abs{d(\nu)}^2}
\biggl\{\biggl[\frac 12 \,G^2\Delta-\nu\,\frac{\cgam \mgam
}2+\left(\om+\nu\right)
\\ {}\times\left(\nu^2- \Delta^{2}-\frac {\cgamq }4\right)\biggr] \left(N(\nu)+
\frac 1 2  \right)+ (\nu\to -\nu)\biggr\},
\label{h(nu)th}
\end{eqnarray}
so that the fluctuation contribution can be written as
\begin{equation}
\langle H\rangle_{\rm fl}=\hbar \om \bigl(\mathcal{N}_{\rm rp}+\mathcal{N}_{\rm th}
+ \mathcal{M}_{\rm th}(\Delta)\bigr);
\label{Hhnu}
\end{equation}
by construction we have $\mathcal{N}_{\rm rp}+\mathcal{N}_{\rm th}+
\mathcal{M}_{\rm
  th}(\Delta)\geq 1/2$.
As it appears, the mean energy density cannot be obtained from the knowledge of
$S_q$ alone, but extra terms are present. Moreover, the contribution
proportional to $\mathcal{M}_{\rm
  th}(\Delta)$ can be negative. Depending on the parameter values, the
extra terms can be actually quite small.  It is important to stress that the
given expression for the mean energy of the resonator holds for any temperature
of the phonon bath, including the case of zero temperature.

We further stress that there is not strict energy equipartition. This can be
expected since the mechanical oscillator is coupled to the cavity through its
position and also the counter-rotating terms contribute to the final result. In
the thermal part the lack of equipartition is due to the terms proportional to
$\Delta$, which are present in $S_p^{\rm th}(\nu)$ and not in $S_q^{\rm
th}(\nu)$. In the radiation pressure part the term with $\Omq$ comes from the
position and the one with $\nu^2$ comes from the momentum and give different
contributions to the mean energy.

\subsubsection{Vanishing effective detuning.}\label{sec:D=0}
For a vanishing effective detuning $\Delta=0$ all the computations can be
performed analytically. The second thermal contribution $\mathcal{M}_{\rm
th}(\Delta)$ vanishes and the coupling constant takes the value $G^2=8\hbar
g_0^{\,2}E^2/\left(m\om \cgamq\right)$. For the spectra of the fluctuations the
explicit expressions reduce to
\begin{equation}\label{Sq0rp-Sq0th}
\eqalign{S_q^{\rm rp}(\nu)&= \frac{\Om \om G^2\cgam}{2\left(\nu^2+\frac{\cgamq }4
\right)\left[\left(\nu-\om\right)^2+\frac \mgamq
4\right]\left[\left(\nu+\om\right)^2+\frac \mgamq 4\right]},
\\
S_q^{\rm th}(\nu)&=\frac{\Om\mgam}{2\om}\biggl[\frac{N(\nu)+\frac 1 2 }
{\left(\nu-\om\right)^2+\frac\mgamq 4}+(\nu\to -\nu)\biggr],}
\end{equation}
leading upon integration to
\begin{eqnarray*}
\frac 1 2\, m \Omq\left(\langle q^2\rangle_{\rm eq}-\langle q\rangle_{\rm
eq}^{\;2}\right)=\frac{\hbar\Omq}{4\om}\left(2N_{\rm
eff}+1\right)+\frac{\hbar\om G^2\left(2\mgam + \cgam  \right)
}{8\mgam \left(\frac{\left(\mgam +\cgam \right)^2}4+\omq\right)},
\\
\frac 1 {2m}\,\langle p^2\rangle_{\rm
eq}=\frac{\hbar\Omq}{4\om}\left(2N_{\rm eff}+1\right)+\frac{\hbar
\om G^2\cgam  }{8\mgam \left(\frac{\left(\mgam +\cgam\right)^2}4+\omq\right)},
\\
\frac \mgam  4\,\langle \left\{q,p\right\}\rangle_{\rm eq}=
-\frac{\hbar\mgamq }{8\om}\left(2N_{\rm eff}+1\right),
\end{eqnarray*}
with $N_{\rm eff}$ as in \eref{eq:7}. These expressions show that equipartition
of the mean energy is not valid just due to the radiation pressure
contributions. However equipartition approximately holds for $\cgam \gg 2\mgam
$, which is the case typically considered in many theoretical studies and
experiments. We further have for the fluctuation contributions to the mean
energy
\[ \fl
\mathcal{N}_{\rm rp}=\frac{G^2\left(\mgam + \cgam  \right)}{4\mgam
  \left(\frac{\left(\mgam +\cgam \right)^2}4+\omq\right)},
\qquad
  \mathcal{N}_{\rm th}=N_{\rm eff}+\frac 12,
\qquad
\mathcal{M}_{\rm th}(\Delta)=0.
\]
The mean equilibrium energy of the mechanical oscillator is thus increased due
to the interaction with the cavity as a consequence of the presence of the
strong laser in resonance. For the values considered in \Fref{fig:plot1} we
have $\mathcal{N}_{\rm rp}\simeq 1.6 \times 10^4$ corresponding to a
temperature of about 7.9 K.

\subsubsection{Laser cooling.}\label{sec:cool}
As discussed in many papers \cite{PGKBBA06,SDHB12,SN-P13,Chen13,Gigan06}, an
important effect which can be described by the quantum models of cavity
optomechanics is the laser cooling of the mechanical resonator.  Since, as
already discussed, we cannot expect equipartition of the mean mechanical
energy, we cannot speak of temperature in a strict sense.  A natural way to
speak about laser cooling is the comparison of the mean energy  of the
fluctuations of the mechanical oscillator in the presence or the absence of the
stimulating laser (corresponding to $\zeta=0$). So, we have to study the value
of the fluctuation contribution \eref{Hhnu} and to compare it to its value for
$\zeta=0$, which is given by $\langle H\rangle_{\rm fl}\big|_{\zeta=0}=\langle
H_\rmm\rangle_{\rm eq}\big|_{\zeta=0}= \hbar\om \left(N_{\rm eff}+\frac
12\right)$.

To obtain explicit analytical formulae for the mean energy we consider the case
of a constant noise spectrum, that is $N(\nu)=\text{const}=N_{\rm eff}$. To
actually perform the calculations we need the expressions of the zeros of
$d(\nu)$; here we consider the generic case given by the Ansatz \eref{dsolved}.
By lengthy computations the integrals over $\nu$ can be exactly performed,
leading to involved formulae explicitly given in \ref{app:meanEn}. In order to
describe cooling effects the relevant contributions can be written in the form
\begin{equation}\label{maincontrib}
\mathcal{N}_{\rm th}= \frac\mgam {\Gamma_\rmm}\,\mathcal{Q} \left(N_{\rm eff}
+\frac 1 2 \right),\qquad \mathcal{M}_{\rm th}(\Delta)=\frac\mgam {\Gamma_\rmm}\,
\mathcal{K} \left(N_{\rm eff}+\frac 1 2 \right),
\end{equation}
where the quantities $\mathcal{Q}$ and  $\mathcal{K}$ are given in equations
\eref{Q2} and \eref{extracontrib}. The expression for $\mathcal{N}_{\rm rp} $
is given in \eref{rpcontrib}. Note that, while $\mathcal{Q}$ is always
positive, depending on the values of the parameters the quantity $\mathcal{K}$
can be also negative. For a large choice of the parameters $\mathcal{Q}$ turns
out to be close to 1.

In the following figures we describe the effective cooling of the mechanical
oscillator, by considering as a figure of merit the \emph{cooling factor}
\begin{equation}
  \label{eq:1}
\mathcal{C}  = \frac\mgam {\Gamma_\rmm} \left(\mathcal{Q} +\mathcal{K}\right).
\end{equation}
We study two cases, corresponding to the parameter regions for which an exact
or approximate analytic evaluation of the different contributions to the mean
energy has been provided. In both cases mass and bare frequency of the mechanic
oscillator are taken to be $m=2.5 \times 10^{-10}$ kg and $\Om=2\pi \times
10^{7}$ Hz, while the mechanical damping factor is $\gamma_\rmm=2\pi
\times10^2$ Hz. We consider a cavity of length $5 \times 10^{-4}$ m and
resonance frequency $\omega_{\rm c}= 2\pi c /(1064\times 10^{-9})$ Hz, driven
by a laser with a power of $5\times 10^{-2}$ W. For the sake of comparison the
values of the fixed parameters are taken from \cite{Genes07}.

We start by studying the case $\Delta=\om$, studied in \Sref{sec:zeros}, in
which the location of the poles can be evaluated exactly, provided one
distinguishes two regions according to the value of the ratio
$(\cgam-\mgam)^2/4G^2$. No approximation is taken in the expression of the
integrals giving the mean energy. If this ratio is above one, verified for a
cavity damping $\cgam> \bar\gamma_{\rm c}$, where the critical damping
$\bar\gamma_{\rm c}$ is introduced in the comments after \eref{xpm} and
corresponds for the considered parameters to $\bar\gamma_{\rm c} \simeq 4.1
\times 10^7$, the effective damping rates $\Gamma_{\rmm}$ \eref{Gc1} and
$\Gamma_{\rm c}$ \eref{Gm1} are actually distinct, while $\Delta_{\rm
eff}=\omega_{\rm eff}^{\rm m}$ and their expression is given by \eref{Deff1}.
By numerical computations we see that the cooling factor $\Ccal$ is a monotonic
increasing function of the cavity damping rate $\cgam$, and around the starting
point $\bar{\gamma}_{\rm c}$ the cooling factor takes the value $2.9 \times
10^{-5}$. In the complementary region, corresponding to $(\cgam-\mgam)^2/4G^2$
below one, the cooling factor is a decreasing function of the cavity damping
rate, so that the optimal cooling is obtained for $\cgam = \bar{\gamma}_{\rm
c}$. In this region, corresponding to $\cgam < \bar{\gamma}_{\rm c}$, we have
$\Gamma_{\rmm}=\Gamma_{\rm c}$ with value given in \eref{G2}, while the
effective frequencies $\Delta_{\rm eff}$ and $\omega_{\rm eff}^{\rm m}$ are
given by the expressions \eref{G2}. To assess the relevance of the various
contributions in \eref{eq:1} we report the values for $\cgam=\bar\gamma_{\rm
c}$: we have ${\mgam}/{\Gamma_\rmm}\simeq 3.05\times 10^{-5}$,
$\mathcal{Q}\simeq .997$ and $\mathcal{K} \simeq -4.18 \times 10^{-2}$; then,
\eref{eq:1} gives $\Ccal\simeq 2.91\times 10^{-5}$, which is a very strong
cooling factor.

\begin{figure}[ht]
  \begin{center}
    \includegraphics[scale=0.9]{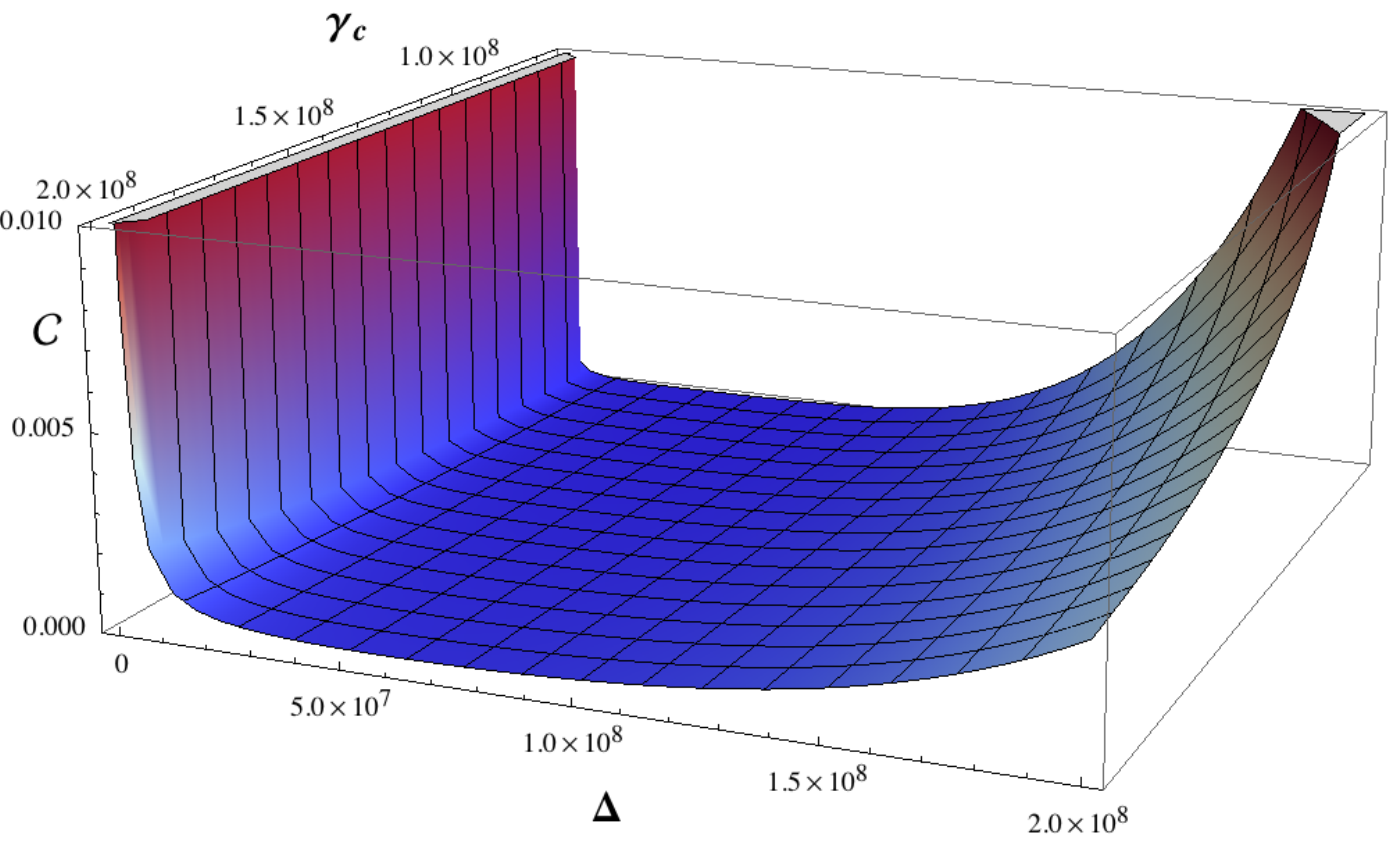}
\caption{Plot of the cooling factor
      $\mathcal{C}$ for the case
      in which the cavity damping is much bigger than the mechanical
      oscillator frequency.  We explore the dependence of the cooling
      factor on both the effective detuning $\Delta$ and the cavity
      damping rate $\cgam$, both expressed in Hz. It appears that the best cooling factor
      is of the order $10^{-3}$ and corresponds to
      $\Delta\lesssim\cgam$.}
    \label{fig:plot1}
  \end{center}
\end{figure}

Instead, in \Fref{fig:plot1} we consider the case $\cgam \gg \om$, that is a
cavity damping much bigger than the mechanical oscillator frequency.
In the exact formulae for the integrals we use the approximate expressions for
$\Gamma_{\rmm}$ and $\Gamma_{\rm c}$ given in \eref{Gammamc}, relying on the
conditions \eref{conditions}. The stationary value of the energy of the
mechanical system has a marked dependence on the effective detuning $\Delta$
and the optimal cooling region, corresponding to $\mathcal{C}$ of the order of
$10^{-3}$, is obtained for $\Delta\lesssim\cgam$.  In this parameter region
$\mathcal{N}_{\rm
  rp}$ can be neglected with respect to $\mathcal{C} N_{\rm eff}$,
unless the phonon bath is below 1 K, so that indeed the quantity $\mathcal{C}$
given in \eref{eq:1} properly describes the cooling effect. When the detuning
$\Delta$ goes to zero the cooling factor rapidly increases in agreement with
the discussion in  \Sref{sec:D=0} showing the presence of heating at
$\Delta=0$; in this parameter region, the cooling effect disappears also when
$\Delta$ grows.

\subsection{Optical spectra}
\label{sec:optical-spectra} We consider now the monitoring of the emitted light
by balanced homodyne and heterodyne detection \cite[Sect.\ 7.2]{BarGreg09}. The
aim is to see which kind of information on the mechanical oscillator can be
obtained by detection of the emitted light.

\subsubsection{Homodyne spectrum.}\label{sec:homo}

The case of a perfect coherent monochromatic local oscillator of frequency
$\omega_0$ with detection of the whole emitted light \cite{BarG08b,BarG13}
corresponds to the continuous measurement of a field quadrature of the type
\begin{equation}
Q(t;\vartheta)=\rmi
\rme^{-\rmi\vartheta}\rme^{\rmi \arg \zeta}
\int_0^t \rme^{-\rmi\omega_0 r}\rmd B_{\rm em}^{\;\dagger} (r)+\text{h.c.};
\end{equation}
$\vartheta$ is a free parameter which depends on the optical path and
determines the observed quadrature. As a consequence of the definition we have
that $[Q(t;\vartheta),Q(s;\vartheta)]=0$. By the properties of the output
fields, discussed after Eq.\ \eref{eq:2}, the commutation rules are preserved;
this gives that the output current $Q^{\rm out}(t;\vartheta):=U(t)^\dagger
Q(t;\vartheta)U(t)$ satisfies $ [Q^{\rm out}(t;\vartheta),Q^{\rm
out}(s;\vartheta)]=0$. This is the key property expressing the fact that
$Q^{\rm out}(t;\vartheta)$ can be measured with continuity in time. Similarly
to \eref{Bnu} we introduce the gated Fourier transforms
\begin{equation}
\fl
Q_T(\nu;\vartheta)=\frac 1 {\sqrt T}\int_0^T\rme^{\rmi\nu t}\rmd
    Q(t;\vartheta),\qquad
\label{ftQout}
Q_T^{\rm out}(\nu;\vartheta)=\frac 1 {\sqrt T}\int_0^T\rme^{\rmi\nu t}\rmd Q^{\rm out}(t;\vartheta).
\end{equation}
From the above relations we obtain the second key relation which guarantees the
presence of commuting observables and therefore the consistency of the theory:
\begin{equation}\label{commut2}
[Q^{\rm out}_T(\nu;\vartheta),Q^{\rm out}_T(\nu';\vartheta)]=0.
\end{equation}

The homodyne spectrum is then given by the expression
\begin{equation} \label{homS}
S(\nu;\vartheta)=\lim_{T\to +\infty}
\Tr\left\{Q^{\rm out}_T(-\nu;\vartheta)Q^{\rm out}_T(\nu;\vartheta)\rho_0\otimes \tilde\envS\right\},
\end{equation}
where the environmental state is given by \eref{tiledeenvS} and $\rho_0$ is any
initial state for the mechanical oscillator and the cavity mode.  Note that
this expression is nothing but the spectrum of the classical stochastic process
representing the output, and not an ad-hoc quantum definition \cite[Sect.\
4]{BarG13}. The commutation property \eref{commut2} implies that \emph{the
homodyne spectrum $S(\nu;\vartheta)$ is an even function of $\nu$}.

As shown in \ref{app:homS}, the homodyne spectrum has both an elastic and an
inelastic component
\begin{equation}\label{S:el+inel}
S(\nu;\vartheta)=S_{\rm el}(\nu;\vartheta)+S_{\rm inel}(\nu;\vartheta),
\end{equation}
which turn out to have the expressions
\begin{equation}\label{Sel+thrp}\eqalign{
S_{\rm el}(\nu;\vartheta)=8\pi\cgam  \abs\zeta^2\left(\sin \vartheta\right)^2 \delta(\nu),
\cr
S_{\rm inel}(\nu;\vartheta)=S_{\rm th}(\nu;\vartheta)+S_{\rm rp}(\nu;\vartheta),}
\end{equation}
with
\begin{equation}\fl \label{Sth}
S_{\rm th}(\nu;\vartheta)=
\frac{2\cgam  \om G^2\left[\left(\frac{\cgam }2\,\cos \vartheta +\Delta
\sin \vartheta\right)^2+\left(\nu\cos\vartheta\right)^2\right]}{\Om
\left(\frac {\cgamq }4+ \left(\Delta -\nu\right)^2\right)\left(\frac {\cgamq }4
+ \left(\Delta +\nu\right)^2\right)}\,S_q^{\mathrm{th}}(\nu),
\end{equation}
\begin{eqnarray}\nonumber\fl
S_{\rm rp}(\nu;\vartheta)
=1+
\frac{2\cgam \om G^2\left[\left(\frac{\cgam }2\,\cos \vartheta +\Delta\sin \vartheta\right)^2
+\left(\nu\cos\vartheta\right)^2\right]}{\Om\left(\frac {\cgamq }4+ \left(\Delta -\nu\right)^2\right)
\left(\frac {\cgamq }4+ \left(\Delta +\nu\right)^2\right)}\,S_q^{\mathrm{rp}}(\nu)
\\ \nonumber{}+\cgam  \om G^2
\RE \Biggl[\frac{\left(\frac {\cgamq }4 +\nu^2-\Delta^{2}\right)\sin 2\vartheta- \Delta\left(\cgam
\cos 2\vartheta-2\rmi\nu\right)}{d(\nu)\left(\frac {\cgamq }4+ \left(\Delta -\nu\right)^2\right)
\left(\frac {\cgamq }4+ \left(\Delta +\nu\right)^2\right)}
\\ \hphantom{anc+4  \cgam  \om G^2nn
}\times \left(\frac{\cgamq }4-\nu^2+\Delta^{2}-\rmi \cgam  \nu\right)\Biggr].\label{Sem}
\end{eqnarray}
Note that all the contributions are indeed positive as shown in \ref{app:homS}.
It is important to stress that the connection between $S_q(\nu)$ and $S_{\rm
inel}(\nu;\vartheta)$ is far from simple. In particular the last contribution
in \eref{Sem} comes from the interference of the electromagnetic part of the
signal with the shot noise, as detailed in \ref{app:homS}. This is a completely
new term, in principle detectable in experiments at very low temperatures.

Let us further stress that different quadratures are incompatible and actually
one can prove the general inequality \cite{BarG08b,BarG13}
\begin{equation}\label{HRineq}
S_{\rm inel}(\nu;\vartheta)S_{\rm inel}(\nu;\vartheta\pm \pi/2)\geq 1,
\end{equation}
which is just a form of the Heisenberg-Robertson uncertainty relations coming
from the canonical commutation relations of the involved Bose fields. As a
result quite different physical information can be extracted from the different
quadratures.

\paragraph{The case $\Delta=0$.} The first striking example of strong
dependence on $\vartheta$ is in the case $\Delta=0$. For $\vartheta = \pi/2$ we
get $S_{\rm el}(\nu;\pi/2)=32\pi E^2  \delta(\nu)/\cgam$ and $S_{\rm
inel}(\nu;\pi/2)=1$: only the shot noise contributes to the inelastic spectrum.

On the contrary, for the quadrature with $\vartheta = 0$ we get $S_{\rm
el}(\nu;\pi/2)=0$ and
\begin{equation}
S_{\rm inel}(\nu;0)=1+\frac{2\cgam  \om G^2}{\Om\left(\frac{\cgamq }4+\nu^2\right)}\,S_q(\nu),
\end{equation}
where $S_q(\nu)$ is now explicitly given by \eref{Sq0rp-Sq0th}. An important
point is that in this case the interference term vanishes exactly and we have a
direct connection of the homodyne spectrum with the fluctuation spectrum of the
position of the mirror. This result has been found also in \cite{GioV01}, but
with the substitution $\hat R(\nu) \to \hat R_{GZ}(\nu)$ in the expression
\eref{Sqth} for $S_q^{\rm th}(\nu)\big|_{\Delta=0}$. As a result, at least in
principle, when $\Delta=0$ the homodyne observation of the quadrature with
$\vartheta=\pi/2$ can give direct experimental information on the correct
expression for $\hat R(\nu)$. At zero temperature one could experimentally
discriminate between our result \eref{T=0,R} and the standard proposal
\eref{T=0,GZ}.

In other cases the interference term does not vanish, but it can be negligible
at high temperatures. For instance, when the interference term is negligible,
at least in the region where $N(\nu)\gg 1$, we recover for $S_{\rm
inel}(\nu;0)$ the result given in \cite[Sect.\ 3]{PGKBBA06}. At high
temperatures the inelastic homodyne spectrum allows to reconstruct the
fluctuation spectrum of position, while no direct information on the
fluctuation of the momentum and on the cross-correlation is obtained. Moreover,
at high temperatures we have also $S_q(\nu)\simeq S_q^{\rm th}(\nu)$; by using
the explicit expressions of $S_q^{\rm th}(\nu)$ \eref{Sqth} and $\hat R(\nu)$
\eref{1.corr.xi} we get
\begin{eqnarray}\nonumber\fl
S_{\rm inel}(\nu;0)\simeq \cgam\mgam  G^2\, \frac{\left(\frac{\cgamq }4 +\nu^2\right)
\left(\frac{\mgamq }4+ \left(\om +\nu\right)^2\right)}{\abs{d(\nu)}^2}
\left(N(\nu)+\frac 1 2 \right) + (\nu \to -\nu).\label{Sth2}
\end{eqnarray}
This expression highlights the dependence of the homodyne spectrum on the
thermal spectrum $N(\nu)$ and the characteristic polynomial $d(\nu)$ \eref{den}
of the dynamical matrix \eref{dynA} of the full optomechanical system.

\paragraph{Squeezing.}
An important information about the non classical nature of the light generated
by optomechanical systems can be obtained considering the quadrature with
$\vartheta=-\pi/ 4$. In the simple case of vanishing detuning $\Delta=0$ and
vanishing temperature $N(\nu)\equiv 0$, it is possible to show from
\eref{S:el+inel}--\eref{Sem} that we have $S_{\rm inel}(0;-\pi/ 4)<1$, at least
in a certain region of the parameters. This means that in a neighbourhood of
$\nu=0$ we have $S_{\rm inel}(\nu;-\pi/ 4)<1$ and the emitted light is
squeezed. This result shows that such a kind of optomechanical systems can
generate non classical light \cite{SDHB12,Pontin14}. Note that, if light
squeezing is present for certain values of the parameters, then the inequality
\eref{HRineq} implies that the complementary quadrature is anti-squeezed. Of
course, experimentally it could be difficult to tune the values of the various
free parameters in order to have squeezing; moreover, the elastic peak in the
spectrum tends to hide the squeezing around $\nu=0$ in the inelastic spectrum.

\subsubsection{Heterodyne spectrum.}\label{sec:hetero}

In the case of heterodyne detection the local oscillator and the stimulating
light are produced by different laser sources; now, the stimulating laser
frequency $\omega_0$ and the local oscillator frequency, say $\mu$, are in
general different. Moreover, the phase difference cannot be maintained stable
and this erases some interference terms. It can be shown \cite{BarPer02},
\cite[Sect.\ 3.5]{Bar06} that the balanced heterodyne detection scheme
corresponds to the measurement in continuous time of the observables
\begin{equation}
I(\mu;t)=\int_0^t \sqrt \varkappa \,\rme^{-\varkappa \left(t-s\right)/2}\,
\rme^{\rmi\mu s+\rmi\alpha}\,\rmd B_{\rm em} (s)+\text{h.c.},
\end{equation}
where $\alpha$ is a phase depending on the optical paths and $\sqrt \varkappa
\,\rme^{-\varkappa t/2}$, \ $\varkappa>0$, represents the detector response
function. As we shall see, the heterodyne spectrum does not depend on $\alpha$.
In the Heisenberg description the observables become the ``output current''
\begin{eqnarray*}\fl
I_{\rm out}(\mu;t)=U(t)^\dagger I(\mu;t)U(t)\\ {}=\sqrt{\varkappa}\int_0^t
\rme^{-\frac \varkappa 2 \left(t-s\right)+\rmi \alpha}\left(\rme^{\rmi\mu s}
\rmd B_{\rm em}(s) + \sqrt\cgam \, \rme^{\rmi\left(\mu-\omega_0\right) s}a_{\rm c}(s)
\rmd s \right) +\text{h.c.}
\end{eqnarray*}
By the definition of $I(\mu;t)$ and the properties of $U(t)$ we get $[I_{\rm
out}(\mu;t),I_{\rm out}(\mu;s)]=0$, which says that the output current at time
$t$ and the current at time $s$ are compatible observables.

While in the homodyne scheme the spectrum of the output is analysed, in the
heterodyne scheme it is usual to register only the output power as a function
of the frequency $\mu$ of the local oscillator. The mean output power of the
detection apparatus at large times is proportional to
\begin{equation}
P(\mu)= \lim_{T\to +\infty}\frac 1 T\int_0^T \rmd t\,\Tr\left\{I_{\rm out}(\mu;t)^2\rho_0\otimes
\tilde\envS\right\};
\end{equation}
the limit is in the sense of the distributions in $\mu$. As a function of
$\mu$, $P(\mu)$ is known as \emph{power spectrum}. Note that to change $\mu$
means to change the frequency of the local oscillator, that is to change the
measuring apparatus. In general $I_{\rm out}(\mu;t)$ and $I_{\rm out}(\mu';s)$
do not commute, even for $t=s$. Then, there is no reason for the power spectrum
to have some symmetry in $\mu$. The heterodyne power spectrum can be decomposed
in an elastic and an inelastic part
\begin{equation}
P(\mu)= \Sigma_{\rm el}(\mu)+ \Sigma_{\rm inel}(\mu),
\end{equation}
\begin{eqnarray}\fl\nonumber
\Sigma_{\rm el}(\mu)= \lim_{T\to +\infty}\frac {\varkappa \cgam } T\int_0^T \rmd t
\left[2\RE\left(\zeta \rme^{\rmi \alpha} \int_0^t \rme^{-\frac \varkappa 2
\left(t-s\right) +\rmi \left(\mu-\omega_0\right)s}\rmd s\right)\right]^2
\\ {}=\frac {\varkappa\cgam  \abs\zeta^2}{\frac{\varkappa^2}4+\left(\mu-\omega_0\right)^2}
\overset{\varkappa\downarrow 0}{ \longrightarrow } 4\pi\cgam  \abs\zeta^2 \,\delta(\mu-\omega_0),
\end{eqnarray}
\begin{equation}\label{Sigmainel}
\Sigma_{\rm inel}(\mu)= \lim_{T\to +\infty}\frac 1 T\int_0^T \rmd t\,\Tr\left\{I_{\rm inel}(\mu;t)^2
\rho_0\otimes \tilde\envS\right\},
\end{equation}
where
\begin{eqnarray*} \fl
I_{\rm inel}(\mu;t)=\sqrt{\varkappa}\int_0^t \rme^{-\frac \varkappa 2 \left(t-s\right)}
\Bigl(\rme^{\rmi\mu s+\rmi \alpha}\rmd B_{\rm em}(s) \\ {}+ \sqrt{\frac\cgam  2}
\, \rme^{\rmi\left(\mu-\omega_0\right) s+\rmi\vartheta}\bigl( Y(s)-\rmi X(s)\bigr)\rmd s \Bigr)
+\text{h.c.}
\end{eqnarray*}

The inelastic part of the spectrum is computed in \ref{app:hetero}. Again it is
possible to identify a radiation pressure contribution and a thermal part
\begin{equation}\label{em+th}
\Sigma_{\rm inel}(\mu)= \Sigma_{\rm rp}(\mu)+\Sigma_{\rm th}(\mu).
\end{equation}
For simplicity we give only the expressions for $\varkappa\downarrow 0$:
\begin{equation}\fl
\Sigma_{\rm rp}(\mu)=1+
\frac{\cgam  \om G^2S_q^{\rm rp}(\mu-\omega_0)}{\Om\left(\frac {\cgamq }4
+\left(\mu-\omega_0-\Delta\right)^2\right)}
-\IM\frac { \cgam  \om G^2}{d(\mu-\omega_0)}\,\frac{\frac\cgam  2
-\rmi\left(\mu-\omega_0+\Delta\right)}{\frac\cgam  2 +\rmi\left(\mu-\omega_0-\Delta\right)}\,,\label{Sigmaem}
\end{equation}
\begin{equation}\label{Sigmath}
\Sigma_{\rm th}(\mu)=
\frac{\cgam  \om G^2S_q^{\rm th}(\mu-\omega_0)}{\Om\left(\frac {\cgamq }4
+\left(\mu-\omega_0-\Delta\right)^2\right)}\,.
\end{equation}
Both contributions are positive as it follows from the expressions
\eref{Sigmath} and \eref{Sigmaem0}. Note the presence of the interference term
in \eref{Sigmaem}. By simple computations one can check that
\begin{equation}
\Sigma_{\rm inel}(\nu+\omega_0)+\Sigma_{\rm inel}(\omega_0-\nu)=S_{\rm inel}
(\nu;\vartheta)+S_{\rm inel}(\nu;\vartheta+\pi/2);
\end{equation}
this is a fundamental relation \cite[Eq.\ (9.61)]{BarGreg09} connecting
heterodyne and homodyne spectra. Moreover, by inserting the definitions of the
relevant quantities given in \eref{1.corr.xi}, \eref{Sqrp} and \eref{Sqth}, an
explicit expression for $\Sigma_{\rm inel}$ can be obtained from which it is
apparent that $\Sigma_{\rm inel}(\mu)>1$: in the heterodyne detection the phase
dependencies are lost and it is impossible to detect squeezing in the emitted
light.

\begin{figure}[h]
  \begin{center}
    \includegraphics[scale=0.9]{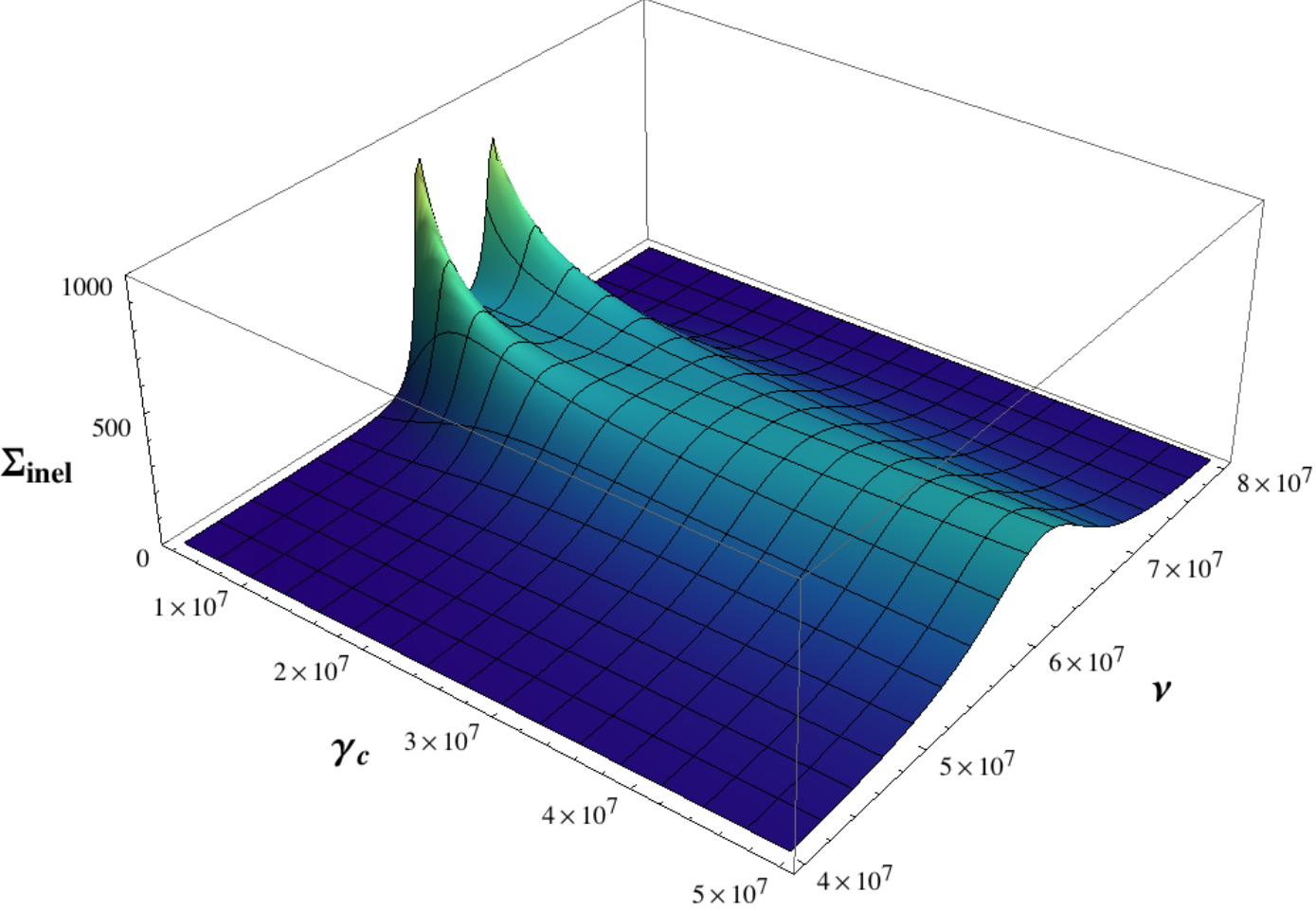}
    \caption{Plot of the inelastic heterodyne spectrum $\Sigma_{\rm inel}$ as a
    function of $\nu$ for a range of values of the cavity damping $\cgam$ around the
    critical value $\bar\gamma_{\rm c}$ discussed in \Sref{sec:cool}. It appears how the two distinct
    peaks of the spectrum coalesce at critical value. The spectrum is plotted for $\Delta=\om$,
    while the other parameters are as in \Sref{sec:cool}.}
    \label{fig:heter}
  \end{center}
\end{figure}

As in the homodyne case, the interference term in \eref{Sigmaem} is negligible
when $N\gg 1$ and we get
\begin{equation}
\Sigma_{\rm inel}(\mu)\simeq 1+
\frac{ \cgam  \om G^2}{\Om\left(\frac {\cgamq }4+\left(\mu-\omega_0-\Delta\right)^2\right)}\,S_q(\mu-\omega_0).
\end{equation}
When this approximation holds, the inelastic heterodyne spectrum too allows to
reconstruct the asymptotic dynamics of the mirror through the position
fluctuations.

To explore the behaviour of the spectrum we take $N(\nu)$ as given by
\eref{N(nu)even} with a Ohmic spectral density. Then, by using the explicit
expressions of $S_q^{\rm rp}$ and $S_q^{\rm th}$ and by setting
$\nu=\mu-\omega_0$, we get
\begin{eqnarray}\nonumber\fl
\Sigma_{\rm inel}(\nu+\omega_0) =1+
\frac{\cgam   G^2}{2\abs{d(\nu)}^2}\biggl\{ \cgam  \omq G^2
\\ {}+ \mgam \left(\frac{\cgamq }4+\left(\nu+\Delta\right)^2\right)
\biggl[\frac{\mgamq }4+\left(\nu-\om\right)^2
+\frac{4\om\abs \nu }{\rme^{\beta \hbar \abs \nu}-1}
\biggr] \biggr\}.
\end{eqnarray}
From this expression we see that the main features of the spectrum will be
determined by the zeros of the denominator $\abs{ d(\nu)}^2$; for instance, as
discussed in \Sref{sec:zeros}, for $\Delta=\om$ we can have one or two
resonance frequencies depending on the value of the cavity decay rate $\cgam$.
In \Fref{fig:heter} we show this phenomenon: the two distinct peaks coalesce as
$\cgam$ increases. For these values of the parameters one can check that the
main contribution to the inelastic heterodyne spectrum comes from the thermal
part $\Sigma_{\rm th}$. It can be checked that in this parameter region the
behaviour of the inelastic homodyne spectrum  $S_{\rm inel}(\nu;0)$ is very
close to the heterodyne one as depicted in \Fref{fig:heter}. Let us notice that
the behaviour shown in \Fref{fig:heter} does not uncover the whole rich
structure of the spectrum which appears by exploring other parameter regions.

\section{Summary and outlook} \label{sec:concl}

In this article we have shown how to give a fully quantum description of a
dissipative mechanical oscillator. The combined use of master equations and
quantum Langevin equations allows for the construction of a dissipative
dynamics respecting symmetries and physical constraints, such as the energy
equipartition at equilibrium, and subject to dissipation with an arbitrary
noise spectrum.  A crucial feature allowing for these results is that for a
mechanical oscillator the definition of the creation and annihilation operators
$a_\rmm$ and $a_\rmm^{\,\dagger}$ in terms of position and momentum is not the
usual one, but rather depends on the damping constant $\mgam $, as discussed in
\Sref{finalME}; the standard result is only recovered for a vanishing damping
constant as can be seen from \eref{atau}.  Moreover, the quantum Langevin
equations for the system, and the input-output relations for the noises, for
both the mechanical oscillator and for the optomechanical system, given in
\Sref{sec:qL} and \Sref{sec:optomechanical-model} respectively, need not be
postulated: they are nothing but the Heisenberg equations of motion determined
by the Hudson-Parthasarathy unitary evolutions \eref{eq:U} and \eref{eq:Umc}.
In this framework it appears that, in order to preserve the Heisenberg
uncertainty relations, the momentum operator can be interpreted as the time
derivative of the position operator only in a ``coarse grained'' picture. An
help in comparing our approach to others and in discussing the structure of the
noises comes from the quantum Langevin equation in Newton form, see
\Sref{sec:Newt}, which at the price of introducing singular noises does not
contain the momentum operator. Indeed in the quantum case important constraints
on the correlation functions of the operator noises come from the fact that
they need to be positive definite and compatible with the commutation rules of
such noises. In this formalism, we are further able to introduce a field analog
of the $P$-representation for the state of the environment and this opens the
possibility of treating an arbitrary noise spectrum as done in \Sref{sec:n-M}.

Our description of the mechanical oscillator is not very different from other
proposals at medium and high temperatures of the phonon bath. Differences
become relevant for very small temperatures. Indeed the dynamics we have
constructed is fully ``quantum'' at all temperatures and this opens the
possibility of constructing models of optomechanical systems which are reliable
also in a deep quantum regime. As an example we have studied a prototypical
system: a mechanical resonator interacting via radiation pressure with a single
optical mode in a cavity. For this case we have given explicit general formulae
for the fluctuation spectra of position and momentum of the mechanical
resonator and for the mean mechanical energy at equilibrium. By using detection
theory in continuous time, we have obtained the full expressions of the
homodyne and heterodyne spectra of the emitted light.  For not too low
temperatures, usual results are recovered, such as laser cooling and connection
between the light spectra and the fluctuations of position of the mechanical
component. However, our description is valid also at very low temperatures,
when semi-classical reasoning is not valid and the observation of the spectra
of the emitted light is not giving a direct measurement of the mechanical
fluctuations.

Many generalizations are possible \cite{MaP11,XBP12,ABV12,HofH14}, which could
benefit of a systematic and consistent treatment. The simplest generalization
is to include imperfections in the detection scheme and noise in the
stimulating laser light \cite{SN-P13,BarPer02,Pontin14,Bar06}. But also direct
detection can be included \cite{Bar06} or the entanglement between resonator
and optical mode can be studied. Moreover, the whole theory has in some sense
``modular'' properties and can be applied to more complicated systems, say when
more mechanical resonators and more optical modes are involved.

\ack B V acknowledges support by EU through the Collaborative Projects QuProCS
(Grant Agreement 641277), by the COST Action MP1006 Fundamental Problems in
Quantum Physics and by UniMI through the H2020 Transition Grant
14-6-3008000-623.

\appendix

\section{{}\qquad\quad \ \ Computation of the fluctuations}\label{app:flu}

The Fourier transformed equations of motion corresponding to \eref{eq:x} can be
solved by purely algebraic manipulations; essentially the problem reduces to
compute the inverse of the matrix $A+\rmi \nu \openone$. By using the
characteristic polynomial $d(\nu)$ of the dynamical matrix $A$ \eref{den} the
final result for large $T$ turns out to be
\begin{equation}\label{sol:F1}\fl
F_1(T;\nu)\simeq \frac {\Delta^{2}+\left(\rmi \nu -\frac \cgam  2 \right)^2}{d(\nu)}
\,\sqrt{\frac{\mgam \Om }{2 \om }}\,Z_1^{\rm th}(T;\nu)+\frac {G }{d(\nu)}\,
\sqrt{\frac{\om\Om\cgam}2}\,Z_1^{\rm rp}(T;\nu),
\end{equation}
\begin{equation}
\label{sol:F2}\fl
F_2(T;\nu)\simeq \frac{\sqrt{\mgam }}{d(\nu)\sqrt{2\om\Om}}\,Z_2^{\rm th}(T;\nu)
- \frac {\rmi \nu G }{d(\nu)}\,\sqrt{\frac{\om\cgam}{2\Om}}\,Z_1^{\rm rp}(T;\nu),
\end{equation}
\begin{equation}\fl
F_3(T;\nu)\simeq\frac {G\Delta}{d(\nu)}\,\sqrt{\frac\mgam 2}\,Z_1^{\rm th}(T;\nu)
+\frac{\rmi\nu\left(\rmi\nu-\mgam  \right)+\Omq}{d(\nu)}\,\sqrt{\frac\cgam 2}
\,Z_1^{\rm rp}(T;\nu),\label{sol:Z1}
\end{equation}
\begin{eqnarray}\nonumber\fl
F_4(T;\nu)\simeq-\frac{\left(\rmi \nu-\frac\cgam 2\right)G}{d(\nu)}\,\sqrt{\frac\mgam 2}\,Z_1^{\rm th}(T;\nu)
+
\frac{\rmi\nu\left(\rmi\nu-\mgam  \right)+\Omq}{d(\nu)}\, \sqrt{\frac \cgam 2}\,Z_2^{\rm rp}(T;\nu)
\\ {}-\frac{G^2\om}{d(\nu)}\, \sqrt{\frac\cgam 2}\left(\frac{ \zeta}{\abs\zeta}\,\hat B_{\rm em}^T(\omega_0-\nu)^\dagger
+\frac{\overline \zeta}{\abs\zeta}\, \hat B_{\rm em}^T (\nu+\omega_0)\right), \label{sol:Z2}
\end{eqnarray}
\begin{equation}\fl
Z_1^{\rm th}(T;\nu)= \left(\rmi\left( \nu+\om\right)-\frac\mgam  2 \right)
\overline \tau\,\hat B_{\rm th}^T(\nu) +\left(\rmi\left( \nu-\om\right)-\frac\mgam  2 \right)
\tau\hat B_{\rm th}^T(-\nu)^\dagger ,
\end{equation}
\begin{eqnarray}\nonumber\fl
Z_2^{\rm th}(T;\nu)&= \biggl[\left(\Delta^{2}+\left(\frac \cgam  2 -\rmi
\nu\right)^2\right) \left(\Omq-\nu \left( \om+\rmi\,\frac\mgam  2
\right)\right) - G^2\om\Delta\biggr] \tau\hat B_{\rm th}^T(-\nu)^\dagger
\\ \fl{}&+
\biggl[\left(\Delta^{2}+\left(\frac \cgam  2 -\rmi \nu\right)^2\right)
\left(\Omq+\nu \left( \om-\rmi\,\frac\mgam  2 \right)\right)-
G^2\om\Delta\biggr]\overline \tau\,\hat B_{\rm th}^T(\nu) ,
\end{eqnarray}
\begin{eqnarray}\nonumber\fl
Z_1^{\rm rp}(T;\nu)=
\left(\rmi\left(\nu-\Delta\right)-\frac \cgam  2 \right)\frac{\zeta}{\abs\zeta}
\,\hat B_{\rm em}^T(\omega_0-\nu)^\dagger
\\  {}+\left(\rmi\left(\nu+\Delta\right)-\frac \cgam  2 \right) \frac{\overline\zeta}{\abs\zeta}
\,\hat B_{\rm em}^T(\nu+\omega_0),
\end{eqnarray}
\begin{equation}\fl
Z_2^{\rm rp}(T;\nu)=
\left(\Delta-\nu-\rmi\,\frac \cgam  2 \right)\frac{\zeta}{\abs\zeta}
\,\hat B_{\rm em}^T(\omega_0-\nu)^\dagger
+\left(\Delta+\nu+\rmi\,\frac \cgam  2 \right) \frac{\overline\zeta}{\abs\zeta}
\,\hat B_{\rm em}^T(\nu+\omega_0).
\end{equation}

To compute the spectra \eref{flsp1}, \eref{flsp2} we need also the field
correlations. From the correlations \eref{BBcorr} and the fact that $B_{\rm
em}$ is a Fock field in the vacuum state we get
\begin{equation}\label{Enu}\fl\eqalign{
\langle \hat B^T_{\rm em}(\nu)\hat B^T_{\rm em}(\nu)^\dagger\ranenv=1, \qquad
\langle \hat B^T_{\rm th}(\nu)^\dagger\hat B^T_{\rm th}(\nu)\ranenv=N(\nu),
\cr
\langle \hat B^T_{\rm em}(\nu)^\dagger\hat B^T_{\rm em}(\nu)\ranenv=0, \qquad
\langle \hat B^T_{\rm th}(\nu)\hat B^T_{\rm th}(\nu)^\dagger\ranenv=N(\nu)+1,}
\end{equation}
while the cross-correlations involving both $B_{\rm th}$ and $B_{\rm em}$
vanish. Also the fluctuations of the cavity mode operators and the correlations
oscillator/mode could be computed by using all the components
\eref{sol:F1}--\eref{sol:Z2}, but we do not study them in this work.

\subsection{{}\qquad\quad \ \ The zeros of $d(\nu)$}\label{app:zeros}

Let us assume that $d(\nu)$ has two zeros of the form $\nu_\rmm= \omega_{\rm
eff}^{\;\rmm}-\rmi \Gamma_\rmm/2$ and $\nu_\rmm= \Delta_{\rm eff}-\rmi
\Gamma_{\rm c}/2$ with $\omega_{\rm eff}^{\;\rmm}\neq 0$ and $\Delta_{\rm
eff}\neq 0$; by the property $\overline{d(\nu)}=d(-\overline \nu)$, the other
two zeros are $-\overline{\nu_\rmm}$ and $-\overline{\nu_{\rm c}}$. Therefore,
we can write $d(\nu)$ in the form \eref{dsolved} or
\begin{equation}\label{dsolved2}
d(\nu)= \left(\nu-\nu_\rmm\right)\left(\nu+\overline{\nu_\rmm}\right)\left(\nu-\nu_{\rm c}\right)
\left(\nu+\overline{\nu_{\rm c}}\right).
\end{equation}
By equating this expression to \eref{den} we get the algebraic system
\begin{equation}\label{system}
\cases{
\Gamma_\rmm+ \Gamma_{\rm c}=\cgam+\mgam,
\cr
\Gamma_{\rm c}\abs{\nu_\rmm}^2+\Gamma_\rmm\abs{\nu_{\rm c}}^2=\cgam\Omq+ \mgam\left(\Delta^2+\frac{\cgamq}4\right) ,
\cr
\abs{\nu_\rmm}^2+\abs{\nu_{\rm c}}^2+\Gamma_\rmm \Gamma_{\rm c}=\Omq+ \Delta^2+\frac{\cgamq}4+\cgam \mgam,
\cr \abs{\nu_\rmm}^2\abs{\nu_{\rm c}}^2=\Omq\left(\Delta^2+\frac{\cgamq}4\right)-G^2\om\Delta.}
\end{equation}
The stability conditions \eref{HRcond1}, \eref{HRcond2} guarantee
$\Gamma_\rmm>0$ and $\Gamma_{\rm c}>0$. By assuming $\Gamma_{\rm c}\neq
\Gamma_{\rmm}$, from this system we get in particular
\begin{equation}\label{moduli}\eqalign{
\Delta_{\rm eff}^{\;2}=\frac{\Gamma_{\rm c}-\mgam}{\Gamma_{\rm c}-\Gamma_{\rmm}}
\,\Delta^2-\frac{\cgam -\Gamma_{\rm c}}{\Gamma_{\rm c}-\Gamma_{\rmm}}\, \omq -
\left(\cgam-\Gamma_{\rm c}\right)\frac{\Gamma_{\rm c}-\mgam}4,
\cr
\omega_{\rm eff}^{\rm m \;2}=\frac{\Gamma_{\rm c}-\mgam}{\Gamma_{\rm c}-\Gamma_{\rmm}}
\,\omq-\frac{\cgam -\Gamma_{\rm c}}{\Gamma_{\rm c}-\Gamma_{\rmm}}\, \Delta^2
-\left(\cgam-\Gamma_{\rm c}\right)\frac{\Gamma_{\rm c}-\mgam}4.}
\end{equation}

\paragraph{The case $\Delta=\om$.} An exact expression for $\Gamma_\rmm$ and
$\Gamma_{\rm c}$ can be found when $\Delta=\om$. We study only the case of
$d(\nu)$ of the form \eref{dsolved2} with four distinct zeros.

In the case $\Gamma_{\rm c} \neq \Gamma_\rmm$ we set $x=\Gamma_{\rm c}
-\Gamma_\rmm$ and insert \eref{moduli} and $\Gamma_{\rm c}
+\Gamma_\rmm=\cgam+\mgam$ into the last equation of the system \eref{system};
in such a way we get
\[
x^4+\left[16\omq-(\cgam-\mgam)^2\right]x^2+ 64 G^2\omq -16\omq (\cgam-\mgam)^2=0.
\]
Then, by using the solution of the equation for $x^2$ and  Eqs.\ \eref{moduli},
we find the result \eref{Gc1}--\eref{u+e}. By imposing $\Gamma_{\rm c}$,
$\Gamma_{\rmm}$, $\Delta_{\rm eff}^{\;2}$ to be real and strictly positive and
$\Gamma_{\rm c}\neq \Gamma_{\rmm}$, we get the necessary and sufficient
condition \eref{condition2x}.

By the choice $\Gamma_{\rm c } = \Gamma_{\rmm}$, from the system \eref{system}
we get directly the result \eref{G2}, \eref{xpm}, together with the conditions
\eref{condition3x}.

\paragraph{An approximate expression.}
To compute approximately $\Gamma_\rmm$ we adapt a suggestion given in
\cite{Genes07,SN-P13} and based on an approximation of the mechanical
susceptibility. In the expression of $d(\nu_\rmm)$ taken from \eref{den} we
make the approximation $\left(\nu_\rmm+\Delta+\rmi\frac \cgam
2\right)\left(\nu_\rmm-\Delta+\rmi\frac \cgam  2\right)\simeq
\left(\om+\Delta+\rmi\frac {\cgam-\mgam} 2\right)\left(\om-\Delta+\rmi\frac
{\cgam-\mgam}  2\right)$ and we solve $d(\nu_\rmm)=0$ for $\Gamma_\rmm$ under
the conditions \eref{conditions}, \eref{chi}. By using also the first equation
of the system \eref{system} we get the expression \eref{Gammamc} for the
damping constants. Once we have $\Gamma_{\rmm}$ and  $\Gamma_{\rm c}$, we can
compute $\omega_{\rm eff}^{\rm m \;2}$ and $\Delta_{\rm eff}^{\;2}$ from the
equations \eref{moduli}, which do not contain approximations.

For consistency we need the positivity of $\Delta_{\rm eff}^{\;2}$ and
$\omega_{\rm eff}^{\rm m \;2}$, which means the positivity of the right hand
sides of equations \eref{moduli}. Under the approximations
\eref{conditions}--\eref{Gammamc}, this gives
\begin{equation}\label{conditions3}\eqalign{
\frac{\Delta^2}{\omq}\gtrsim \frac{\chi(\Delta)}{1-\chi(\Delta)}
+\chi(\Delta)\bigl(1-2\chi(\Delta)\bigr)\frac{\left(\cgam- \mgam\right)^2}{4\omq},
\cr
\chi(\Delta)\Biggl[\frac{\Delta^2}{\omq}+\bigl(1-\chi(\Delta)\bigr)
\bigl(1-2\chi(\Delta)\bigr)\frac{\left(\cgam- \mgam\right)^2}{4\omq}\biggr]
\lesssim 1-\chi(\Delta);}
\end{equation}
because $\chi(\Delta)$ has the same sign as $\Delta$, conditions
\eref{conditions3} give true restrictions only for $\Delta>0$. We see also that
conditions \eref{conditions3} are violated for $\Delta $ positive and small. In
this situation the cavity is overdamped and the decomposition of $d(\nu)$ takes
the form $d(\nu)=\left(\nu - \omega_{\rm eff}^{\rm m}+\rmi
\,\frac{\Gamma_\rmm}2\right)\left(\nu + \omega_{\rm eff}^{\rm m}+\rmi
\,\frac{\Gamma_\rmm}2\right)\left(\nu+\rmi \,\frac
{\Gamma_1}2\right)\left(\nu+\rmi \,\frac {\Gamma_2}2\right)$; we do not study
this case.

\subsection{{}\qquad\quad \ \ Computation of the mean mechanical energy}\label{app:meanEn}

The integrals over $\nu$ in \eref{Nrp}, \eref{h(nu)th} can be performed by the
residue method, under the Ansatz \eref{dsolved} and $N(\nu)\equiv N_{\rm eff}$.
First we set
\begin{equation}\eqalign{
D^2=\biggl(\Delta_{\rm eff}^{\;2}+\omega_{\rm eff}^{\rm m \;2}+\frac{\left(\cgam+\mgam\right)^2}4 \biggr)^2
-4\omega_{\rm eff}^{\rm m \;2}\Delta_{\rm eff}^{\;2},
\cr
L_{\pm}=\frac{\cgamq \mp \Gamma_{\rm c}^{\,2}}4-\Delta^2 \pm \Delta_{\rm eff}^{\;2},
\qquad \Omega_{\rm eff}^{\rm m \;2}=\omega_{\rm eff}^{\rm m \;2}+\Gamma_\rmm^{\,2}/4. }
\end{equation}
With this notation we have
\begin{eqnarray}\nonumber\fl
\mathcal{N}_{\rm rp}=  \frac{G^2 \cgam}{4\Gamma_\rmm\Gamma_{\rm c}D^2}
\biggl\{ \frac{G^2\om \Delta}{2\abs{\nu_{\rm c}}^2\abs{\nu_\rmm}^2}
\left[\mgam\Omq+\cgam\left( \Delta^2 +\frac{\cgamq}4\right)
+\mgam\cgam\left(\mgam+\cgam\right)\right]
\\ {}+\biggl( \Delta^2+\omq +\frac{\left(\cgam+\mgam\right)^2}4\biggr)
\left(\cgam+\mgam\right)\biggr\}, \label{rpcontrib}
\end{eqnarray}
where $\abs{\nu_{\rm c}}^2\abs {\nu_{\rmm}}^2$ is given by the last of
\eref{system}. The thermal contributions $\mathcal{N}_{\rm th}$ and
$\mathcal{M}_{\rm th}(\Delta)$ are given in \eref{maincontrib} in terms of the
expressions
\begin{eqnarray}\nonumber\fl
\mathcal{Q}=\frac{\Omq +\Omega_{\rm eff}^{\rm m \;2}}{2\Omega_{\rm eff}^{\rm m \;2}}+\frac{L_{+}}{2\Gamma_{\rm c}D^2}\biggl\{
\left(\cgam+\mgam\right)
\frac{L_{-}+2\Omq}{16}
+2 \cgam\Omq+ 2\mgam\left(\Delta^2+\frac\cgamq 4 \right)
\\  {}+\frac{\Omq L_{-}}{\abs {\nu_{\rm c}}^2\abs {\nu_{\rmm}}^2}
\left[ \cgam\left(\Delta^2+\frac\cgamq 4\right)+\mgam \Omq+\cgam \mgam\left(\cgam+\mgam\right)\right]\biggr\}
, \label{Q2}
\end{eqnarray}
\begin{eqnarray}\nonumber\fl
\mathcal{K}=\frac{G^2\Delta}{2\Gamma_{\rm c} D^2}
\biggr\{\frac{\left(\Delta^2+\frac{\cgamq}4\right)\left(\frac\mgamq 4
- \omq\right)}{2\om \abs{\nu_{\rm c}}^2\abs{\nu_\rmm}^2}\left[\mgam\Omq
+\cgam\left(\Delta^2+\frac{\cgamq}4\right)+\cgam\mgam\left(\cgam+\mgam\right)\right]
\\ {}+\om\left(\frac \mgam 2 +\cgam\right) -\frac{\cgam\left(\Delta^2+\frac\cgamq 4\right)
+\mgam\left(\frac\mgam 2 +\cgam \right)^2}{2\om}\biggr\}. \label{extracontrib}
\end{eqnarray}
Note that, while $\mathcal{Q}$ is always positive, $\mathcal{K}$ can also take
on negative values.

\section{{}\qquad\quad \ \ Computation of the optical spectra}
\subsection{{}\qquad\quad \ \ The homodyne spectrum}\label{app:homS} The homodyne spectrum
\eref{homS} involves the quantity $Q^{\rm out}_T(\nu;\vartheta)$ \eref{ftQout};
by the rules of quantum stochastic calculus we can compute $\rmd Q_T^{\rm
out}(\nu;\vartheta)$, which turns out to contain the quantities \eref{sol:Z1},
\eref{sol:Z2}. Then, by integration we obtain
\begin{equation}\label{QoutApp}\fl
Q^{\rm out}_T(\nu;\vartheta)\simeq 4\sqrt\cgam  \abs\zeta\sin \vartheta\; \rme^{\rmi \nu T/2}
\frac{\sin \nu T/2}{\nu \sqrt{ T}}+Q^{\rm th}_T(\nu;\vartheta)+Q^{\rm em}_T(\nu;\vartheta),
\end{equation}
\[
Q^{\rm th}_T(\nu;\vartheta)=\overline{E_{\rm th}(\nu;\vartheta)\tau}\,
\hat B_{\rm th}^T(\nu)+E_{\rm th}(-\nu;\vartheta) \tau\hat B_{\rm th}^T(-\nu)^\dagger,
\]
\begin{eqnarray*}\fl
Q^{\rm em}_T(\nu;\vartheta)=-\overline{E_{\rm em}(\nu;\vartheta)}\, \rmi\rme^{\rmi\left(\vartheta-
\arg \zeta\right)}\hat B_{\rm em}^T(\nu+\omega_0)\\ {}+E_{\rm em}(-\nu;\vartheta)
\rmi\rme^{-\rmi\left(\vartheta-\arg \zeta\right)} \hat B_{\rm em}^T(\omega_0-\nu)^\dagger,
\end{eqnarray*}
\begin{equation}
E_{\rm th}(\nu;\vartheta)=-G
\sqrt{\mgam  \cgam }\left(\frac\mgam  2 +\rmi\left( \nu+\om\right)\right) L(\nu;\vartheta),
\end{equation}
\begin{equation}
E_{\rm em}(\nu;\vartheta)=- \frac {\frac \cgam  2-\rmi \left(\nu-\Delta \right)}{\frac \cgam  2
+\rmi \left(\nu-\Delta \right)} +\frac{\rmi \om \cgam G^2\rme^{\rmi \vartheta}L(\nu;\vartheta)}
{\frac \cgam  2+\rmi \left(\nu-\Delta \right)}\label{F(nu)},
\end{equation}
\begin{equation}
L(\nu;\vartheta)=\frac{\Delta\sin \vartheta+\left(\frac \cgam  2 +\rmi \nu\right)\cos \vartheta }
{d(-\nu)}.
\end{equation}
Note that $ L(-\nu;\vartheta)=\overline{L(\nu;\vartheta)}$. The key relation
\eref{commut2} together with $[\hat B_i^T(\nu),\, \hat B_i^T(\nu)^\dagger]=1$
implies
\[
[Q^{\rm th}_T(\nu;\vartheta)+Q^{\rm em}_T(\nu;\vartheta),\, Q^{\rm th}_T(-\nu;\vartheta)
+Q^{\rm em}_T(-\nu;\vartheta)]=0,
\]
which is equivalent to
\begin{equation}\label{EF0}\fl
\abs{ E_{\rm th}(\nu;\vartheta)}^2-\abs {E_{\rm th}(-\nu;\vartheta)}^2
+\abs {E_{\rm em}(\nu;\vartheta)}^2-\abs {E_{\rm em}(-\nu;\vartheta)}^2=0.
\end{equation}
By long computations this relation can be verified also explicitly by using the
expressions of $E_{\rm th}(\nu;\vartheta)$ and $E_{\rm em}(\nu;\vartheta)$.

By using \eref{QoutApp}, \eref{EF0} and \eref{Enu}, from \eref{homS} we get the
decomposition of the homodyne spectrum expressed by Eqs.\ \eref{S:el+inel},
\eref{Sel+thrp} with
\[
S_{\rm th}(\nu;\vartheta)=\abs{E_{\rm th}(\nu;\vartheta)}^2\left(N(\nu)+\frac 12\right)
+\abs{E_{\rm th}(-\nu;\vartheta)}^2\left(N(-\nu)+\frac 1 2\right),
\]
\begin{equation}\label{Sem>0}
S_{\rm rp}(\nu;\vartheta)=\frac 12 \left(\abs{E_{\rm em}(\nu;\vartheta)}^2
+\abs{E_{\rm em}(-\nu;\vartheta)}^2\right).
\end{equation}
Note that $S_{\rm  th}(\nu;\vartheta) \geq 0$ and $S_{\rm
rp}(\nu;\vartheta)\geq 0$. To compute the thermal part we note that
$\abs{E_{\rm th}(\nu;\vartheta)}^2$ can be written by using $\hat R(\nu)$
\eref{1.corr.xi}; by taking $S_q^{\mathrm{th}}(\nu)$ from \eref{Sqth}, we get
Eq.\ \eref{Sth}.

To compute the radiation pressure component of the spectrum, we need the square
modulus of $E_{\rm em}$ \eref{F(nu)}, which is the sum of two terms. So, we
have the square modulus of the first term (the shot noise), the square modulus
of the second term (the signal) and the double product (the interference term):
\begin{eqnarray*}\fl
\abs{E_{\rm em}(\nu)}^2=1 +\frac{\omq \cgamq  G^2\abs{L(\nu;\vartheta)}^2}{\frac {\cgamq }4
+ \left(\nu-\Delta \right)^2}
\\ {}+\om  \cgam  G^2  \RE \frac{\rmi \rme^{-2\rmi\vartheta}\left(\frac \cgam  2
-\rmi\left(\nu-\Delta\right)\right) +\rmi \left(\frac \cgam  2-\rmi\left(\nu+\Delta\right)\right)}
{d(\nu)\left(\frac\cgam 2 +\rmi\left(\nu-\Delta\right)\right)}.
\end{eqnarray*}
By inserting this expression into \eref{Sem>0} we get
\begin{eqnarray*}\fl
S_{\rm rp}(\nu;\vartheta)
=1+\frac{\omq \cgamq G^4\left(\frac {\cgamq }4+ \Delta^{2}+\nu^2\right)}{\left(\frac {\cgamq }4
+ \left(\Delta -\nu\right)^2\right)\left(\frac {\cgamq }4+ \left(\Delta +\nu\right)^2\right)}
\abs{\frac{\Delta\sin \vartheta+\left(\frac \cgam  2 +\rmi \nu\right)\cos \vartheta }{d(\nu)}}^2
\\ \fl{}+\om  \cgam  G^2\biggl[
\RE\frac{\rmi \rme^{-2\rmi\vartheta} \left(\frac \cgam 2 -\rmi \left(\nu-\Delta\right)\right)^2
+\rmi \left(\left(\frac \cgam 2 -\rmi \nu\right)^2+\Delta^2\right)} {2d(\nu)\left(\frac \cgamq 4
+ \left(\Delta -\nu\right)^2\right)} +(\nu\to - \nu)\biggr].
\end{eqnarray*}
Finally, by elaborating the argument of the real part and by using the
expression \eref{Sqrp} for $S_q^{\mathrm{rp}}(\nu)$ we get Eq.\ \eref{Sem}.

\subsection{{}\qquad\quad \ \ The heterodyne spectrum}\label{app:hetero} By a procedure similar to
the one used  in \ref{app:flu} and \ref{app:homS}, in the limit of
$\varkappa\downarrow 0$, $\varkappa t \to +\infty$, we get
\begin{eqnarray*}\fl
I_{\rm inel}(\nu;t)\simeq
\rme^{\rmi \alpha}\sqrt \varkappa\int_0^t \rme^{-\frac\varkappa 2 \left(t-s\right)
+\rmi \mu s }\biggl\{\biggl[-\frac{\frac\cgam  2 +\rmi\left(\mu -\omega_0-\Delta\right)}
{\frac\cgam  2 -\rmi\left(\mu -\omega_0-\Delta\right)}-\frac{\rmi\hbar g_0^{\,2}\cgam
\abs\zeta^2}{md(\mu-\omega_0)}
\\ {}\times\frac{\frac\cgam  2 -\rmi\left(\mu -\omega_0+\Delta\right)}{\frac\cgam  2
-\rmi\left(\mu -\omega_0-\Delta\right)}+\frac{\rmi\hbar g_0^{\,2}\cgam {\overline \zeta}^{\,2}}
{md(\omega_0-\mu)}\,\rme^{-2\rmi\left(\mu-\omega_0\right)s-2\rmi \alpha}\biggr]\rmd B_{\rm em}(s)
\\ {}+\rmi \rme^{-\rmi \omega_0s}g_0\overline \tau \sqrt{\frac{\hbar\mgam \cgam }{2m\om}}
\biggl[\frac{\overline \zeta}{d(\omega_0-\mu)}\left(\frac \cgam  2 +\rmi\left(\mu -\omega_0+\Delta\right)\right)
\\ {}\times\left(\frac \mgam  2 +\rmi\left(\mu -\omega_0-\om\right)\right)
\rme^{-2\rmi \left(\mu-\omega_0\right)s-2\rmi \alpha}-\frac{ \zeta}{d(\mu-\omega_0)}
\\ {}\times\left(\frac \cgam  2 -\rmi\left(\mu -\omega_0+\Delta\right)\right)
\left(\frac \mgam  2 -\rmi\left(\mu -\omega_0+\om\right)\right)\biggr]\rmd B_{\rm th}(s)\biggr\}+\text{h.c.}.
\end{eqnarray*}
By using the field correlations \eref{Enu} we can compute the heterodyne
spectrum; again, by the vanishing of the field cross-correlations, the thermal
contribution and the electromagnetic contributions decouple in the expression
of the spectrum. By some long manipulations and by recalling that the limit in
\eref{Sigmainel} is in the sense of distributions, we get Eq.\ \eref{em+th}
with the thermal part given by \eref{Sigmath} and
\begin{equation}\fl\label{Sigmaem0}
\Sigma_{\rm rp}(\mu)=\abs{1+\frac {\rmi \om \cgam  G^2}{2d(\mu-\omega_0)}\frac{\frac \cgam  2
-\rmi \left(\mu-\omega_0+\Delta\right)} {\frac \cgam  2 +\rmi \left(\mu-\omega_0-\Delta\right)}}^2
+
\frac {\omq \cgamq G^4}{4\abs{d(\mu-\omega_0)}^2}\,,
\end{equation}
which becomes \eref{Sigmaem} by expanding the absolute value and using
\eref{Sq0rp-Sq0th}.

\end{document}